\documentclass[12pt,letterpaper]{article}
\usepackage{savesym}
\usepackage{pdflscape}
\usepackage{wrapfig}
\usepackage{amssymb}
\usepackage{amsthm}
\usepackage{amsmath}
\usepackage{graphicx}
\usepackage[usenames, table]{xcolor}
\usepackage{verbatim}
\usepackage{times}
\usepackage[titles]{tocloft}
\usepackage{enumitem}
\usepackage{bm}
\usepackage{sidecap}
\usepackage{epstopdf}
\usepackage{pdfpages}
\usepackage{hyphenat}
\usepackage[colorinlistoftodos]{todonotes}
\usepackage{soul}
\usepackage{multicol}
\usepackage{float}
\usepackage{multibib}
\usepackage{geometry}
\usepackage[small,compact]{titlesec}
\usepackage[normal,labelsep=colon,skip=2pt]{caption}
\usepackage{setspace}
\usepackage{subfig}
\usepackage{tocloft}
\usepackage{url}
\usepackage{floatflt}
\usepackage{array}
\usepackage{wasysym}
\usepackage[round]{natbib}
\usepackage{pgfgantt}
\usepackage[colorlinks=true,
            linkcolor=blue,
            urlcolor=red,
            citecolor=blue]{hyperref}

\usepackage{sidecap}
\graphicspath{{./figs/}{../figs/}{../}}
\savesymbol{comment} 
\usepackage{changes}
\definechangesauthor[color=purple]{xl}
\definechangesauthor[color=orange]{hx}

\definecolor{lightblue}{rgb}{.90,.95,1}
\definecolor{darkgreen}{rgb}{0,.5,0.5}

\let\OLDthebibliography\thebibliography
\renewcommand\thebibliography[1]{
  \OLDthebibliography{#1}
  \setlength{\parskip}{0pt}
  \setlength{\itemsep}{0pt plus 0.3ex}
}

\begin{document}
%%%%%%%%%%%%%%%%%%%%%%
%%%%%%%%%%%%%%%%%%%%%%%%%%%%%%%%%%%%%%%%%%%%%%%%%%%%%%%%%%%%%%%%%%%%%%%%%%%%%%%%%%%%%%%%%

\thispagestyle{empty}
\begin{center}

\Large{\bf Bifurcation and multiple states in plane Couette flow with spanwise rotation}\\
\vspace{2mm}
\normalsize{Xiang Yang$^1$ and Zhen-hua Xia$^2$}\\
$^1$ Mechanical Engineering, Pennsylvania State University, PA, USA, 16802\\
$^2$ Department of Engineering Mechanics, Zhejiang University, Hangzhou, China, 310027   
\end{center}

\begin{abstract}
We present a derivation that begins with the Navier--Stokes equation and ends with a prediction of multiple statistically stable states identical to those observed in a spanwise rotating plane Couette flow.
This derivation is able to explain the presence of multiple states in fully developed turbulence and the selection of one state over the other by differently sized computational domains and different initial conditions.
According to the present derivation, two and only two statistically stable states are possible in an infinitely large plane Couette flow with spanwise rotation, and that multiple states are not possible at very slow or very rapid rotation speeds. 
We also show the existence of limit cycles near statistically stable states. 
\end{abstract}

\section{Introduction}

{%Turbulence is ubiquitous in nature and engineering applications. 
%After more than one hundred years' effort by the community, the classical turbulence theory has been established. 
According to the classical theory, turbulence is often thought to be ergodic when it is fully developed, }
%Once fully developed, turbulence is often thought to be ergodic, 
that is, a turbulent flow visits all viable states in the phase space, leading to a unique statistical state, {which is irrelevant to the initial conditions, at fixed control parameters} \citep{frisch1995turbulence,tsinober2001informal,galanti2004turbulence}. 
This ergodic assumption has profound influence on our thinking and how we approach a turbulent flow.
For example, turbulence scalings, e.g., the logarithmic mean flow scaling {red}{in wall-bounded turbulence} \citep{marusic2013logarithmic} and the -5/3 scaling of the energy spectrum, are defined assuming that the flow has one unique statistical state.

However, ergodicity of fully developed turbulence was challenged by multiple authors in the recent literature {who observed multiple states in von K\'{a}rm\'{a}n flow \citep{ravelet2004multistability, ravelet2008supercritical}, Rayleigh-Benard convection \citep{xi2008flow, ahlers2011heat, vanderpoel2011, weiss2013effect, xie2018}, rotating Rayleigh-B\'enard convection~\citep{Stevens2009,Weiss2010,Wei2015},  Rayleigh-B\'enard convection with tilted containers~\citep{WangQ2018}, Taylor-Couette flow (TCF)~\citep{huisman2014multiple, van2016exploring, gul_2018}, spherical-Couette flow \citep{zimmerman2011bi}, Spanwise rotating plane Couette flow (RPCF)~\citep{xia2018multiple, huang2019hysteresis, Xia2019-analysis}, forced rotating turbulence~\citep{Yokoyama-PRF} and double diffusive convection turbulence~\citep{YangYT-DDC}. In some of the multistate cases, the state transition happens frequently at certain parameter range~\citep{xi2008flow, xie2018}, while in other cases the different states are stable and robust, exhibiting hysteresis behavior when the control parameter varies~\citep{huisman2014multiple, Yokoyama-PRF, huang2019hysteresis, YangYT-DDC}.}
%\citep{ravelet2004multistability, xi2008flow,  cortet2010experimental,van2011connecting, huisman2014multiple, van2016exploring}. 
\cite{huisman2014multiple} showed that different phase-space trajectories could lead to different torques and velocity distributions at the same flow condition in a {TCF at a Reynolds number $Re\sim~10^6$}, suggesting the presence of two possible turbulent states in a laboratory context. {The multiple states in TCF was also reported by~\citet{van2016exploring} who observed multiple states in two experimental setups with different aspect ratios for the working Taylor numbers covering almost two decades, inferring that multiple states in highly turbulent TCF are very robust and they may be expected to persist when $Ta\ge 10^{13}$. \cite{huang2019hysteresis} reported the hysteresis behavior in %{\color{BrickRed}Rotating Taylor Couette Flow} (RPCF) 
RPCF based on two groups of direct numerical simulations at $Re_w=1300$ (see the definition below) with the rotation number $Ro$ varying sequentially in steps in two opposite directions, which is mimicking the experimental setups in \cite{huisman2014multiple}. With fixed streamwise and spanwise lengths of computational domain, $8\pi h$ and $4\pi h$, the flow prefers a state with two pairs of roll cells when $Ro$ increases from 0.02 to 0.5 while it prefers the state with three pairs of roll cells when $Ro$ decreases from 0.5 to 0.02. The flow structures and turbulent statistics also lead to the hysteresis loops.}

These observations have brought many interesting discussions, nonetheless, the research on this topic is, by and large, phenomenological. That is, conclusions were drawn from data with very little or no idea whether they could be applied to a different flow. %; and many questions remain open. 
{In \cite{huisman2014multiple}, they conjectured that the selectability of the large-scale coherent structures plays a very important role in the multiple states of TCF. \citet{van2016exploring} also mentioned that a theoretical understanding of the values of rotation rate at which the systems transits between states remains elusive. Although \cite{Xia2019-analysis} showed the large-scale roll cells at different flow states indeed play an important role through detailed turbulent statistics in RPCF, they have not answered the origin of the two states, and many related questions remain open.}
%Given that multiple states exist in rotating plane Couette flow, one may ask: do multiple states exist in non-rotating plane Couette flow? 
%One can also ask: do multiple states exist in rapidly rotating plane Couette (Taylor-Couette) flow?
%Moreover, many have reported two statistically stable states, but are there more than two possible statistically stable states at one flow condition?
Here, we name a few.
First, it is not clearly whether multiple states exist at all rotation speeds.
Second, it is not known if there could be more than two statistically stable states at a given flow condition.
Other unexplained observations include the selection of one state over the other by differently sized computational domains and the selection of the states by initial conditions \citep{Xia2019-analysis}.
In all, a {theoretical analysis} %theory 
is in need to help to %explain/
{interpret} the experimental and computational observations. 
The objective of this work is to fill in this gap as much as we can.

Before proceeding with our derivation, we briefly review the previous theoretical works. 
From a fundamental standpoint, many of the ideas in the following sections are similar to these expressed in the field of statistical state dynamics (SSD) \citep{farrell2007structure, farrell2014statistical} and the related fields of research \citep{taira2017modal, farrell2017statistical}. 
In particular, SSD recognizes that the dynamics of turbulent statistical state is time dependent, in which case the statistics obtained from many realizations would not in general correspond to a representation of the statistical state at any time.
While this implies ``multiple'' states, SSD typically follows Kolmogorov and assumes ergodicity, i.e., the statistical average of a turbulent system asymptotically approach a fixed point.
Operationally, we will apply truncated Galerkin projection (TGP) of the Navier--Stokes ({NS}) equation to a pre-defined sub-space.
TGP sees most use in dynamical systems \citep{majda2000remarkable,rapun2010reduced}.
A recent application of this methodology in turbulence research could be found \cite{anderson2019non}, which led to non-periodic phase-space trajectories of roughness-driven secondary flows in high Reynolds number boundary layers. 

The rest of the paper is organized as follows.
A more in-depth summary of the multiple states phenomenology in spanwise rotating plane Couette flow is presented in section \ref{sect:data}.
{We will use this data to anchor our derivation.}
We present our derivation in sections \ref{sect:dynamics-data}.
The results are presented in section \ref{sect:results}, followed by conclusions in section \ref{sect:conclusion}.
We try to keep the derivation simple and analytically tractable.
To that end, discussion that involves numerical tools and tediously long algebra are moved to the appendices.

\section{Multiple states phenomenology in plane Couette flow}
\label{sect:data}

\begin{figure}
\centering
\vspace{2mm}
\includegraphics[width=0.7\textwidth]{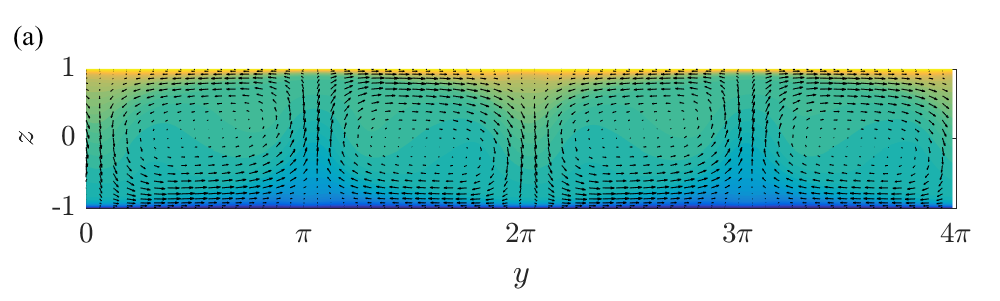}
\includegraphics[width=0.7\textwidth]{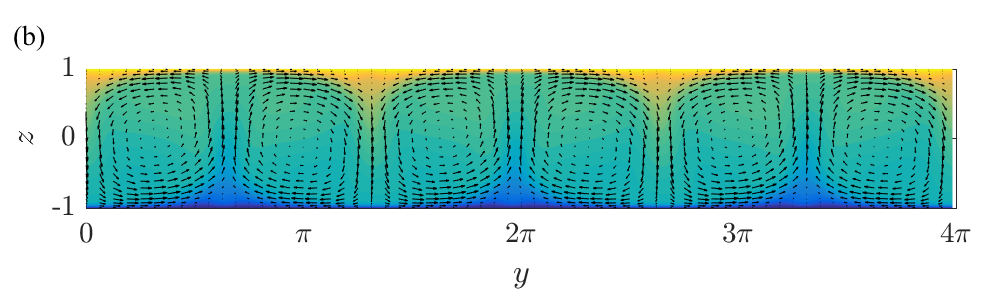}
\caption{(a) Contours of the time and streamwise averaged streamwise velocity.
The velocity shows the in-plane motion.
This statistically stable state features two pairs of roll cells.
(b) Same as (a) but for a state that features three pairs of roll cells.
Here, yellow represents high {streamwise} velocity, blue represents low velocity, and green is in between.
The exact values are not important here.
}
\label{fig:xia-R23}
\end{figure}

%Discussion in this paper will focus on plane Couette flow, as we have access to Xia's data.
In this section, we present a more in-depth summary of the multiple states phenomenology in plane Couette flow. 
Again, we will use these data to anchor our theoretical derivation.
{\cite{xia2018multiple} reported two statistically stable states in spanwise rotating channel at $Re_w=U_wh/\nu=1300$ and $Ro=2\Omega h/U_w=0.2$.
{It may be worth mentioning that the coordinate system in \cite{xia2018multiple} is different from the one here. 
Per the coordinate system here, the spanwise rotation is in the $-y$ direction.}
The two states feature two and three pairs of roll cells in a periodic domain of size $L_x\times L_y\times L_z=10\pi h \times 4\pi h \times 2h$, as shown in figure \ref{fig:xia-R23}. 
Here, $h$ is the half channel height, $U_w$ is half of the velocity difference between two plates, $\Omega$ is the angular velocity in the spanwise direction, $\nu$ is the kinematic viscosity,} $x$, $y$, and $z$ denote the streamwise, spanwise, and wall-normal directions, $L_x$, $L_y$, and $L_z$ are the domain sizes in the $x$, $y$, and $z$ directions. 
Throughout the paper, 
%\textcolor{red}{we will focus our analysis on this domain size with $L_z=4\pi h$, and 
the two states with two pairs and three pairs of roll cells will be referred to as R2 and R3 respectively.
%}

\begin{figure} 
\centering
\vspace{2mm}
\includegraphics[width=0.4\textwidth]{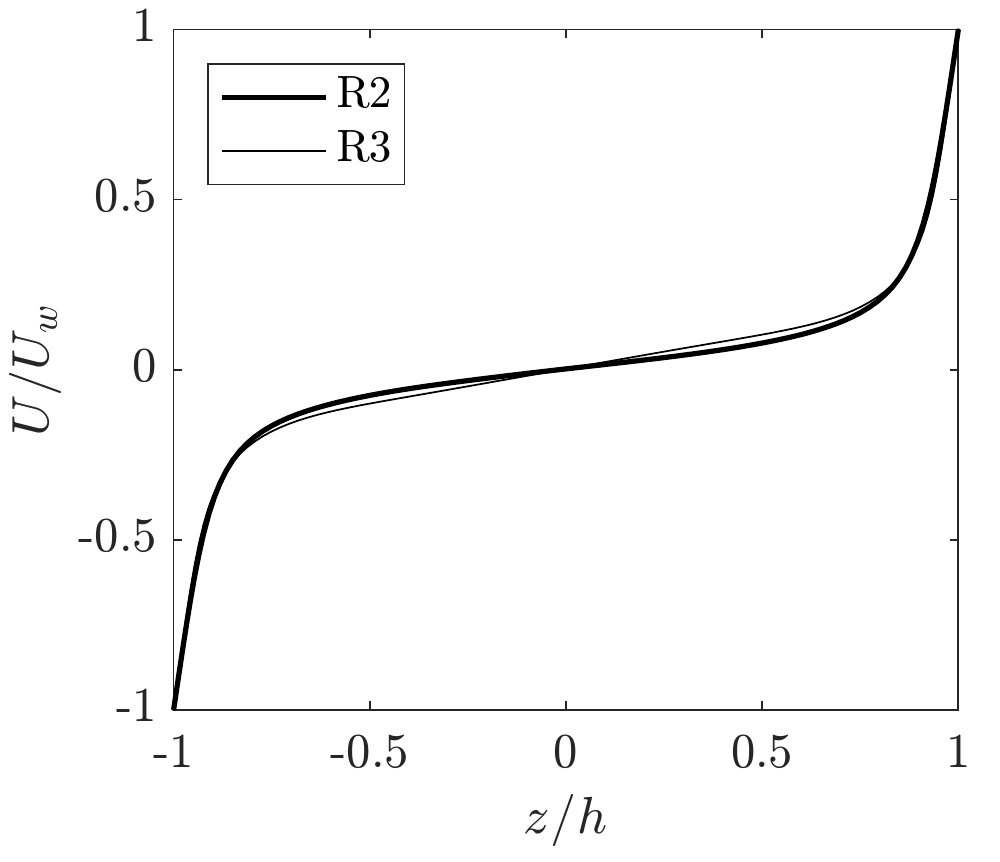}
\caption{Horizontally and time averaged streamwise velocity.
Normalization is by $U_w$ and $h$.
}
\label{fig:xia-um}
\end{figure}

The data in \cite{xia2018multiple} suggests that the roll cells in both R2 and R3 are nearly streamwise homogeneous. 
{This observation will lead to one of the simplifications in the next section.}
Moreover, while R2 and R3 lead to disparate dispersive stresses, the mean flows in R2 and R3 are very similar, see figure \ref{fig:xia-um}.
%\cite{xia2018multiple} also reported two stable states in a $L_z= 8\pi h$ domain.
%There, the two stable states feature 4 and 6 pairs of roll cells.
%Most interestingly, a stable state with 5 pairs of roll cells was \textcolor{red}{not} reported.
In a follow-up work, \cite{huang2019hysteresis} found that the flow is more likely to converge to R2 when increasing the rotation speed from {$Ro=0.02$ to $Ro=0.5$} and the flow is more likely to converge to R3 when decreasing the rotation speed from {$Ro=0.5$ to $Ro=0.02$}.
In another follow-up work, \cite{Xia2019-analysis} synthesized a few initial conditions by linearly interpolating between R2 and R3 and reported that the flow is more likely to converge to R2 if the interpolation is biased towards R2 and vice versa. 
{Furthermore, they showed that the energy transfer from the mean field to the secondary and residual fields are different at the two states, where most of the kinetic energy is transferred to the residual fields through the streamwise secondary flows in R3 state while the transfer directly from the mean field through the production term dominates over other terms in R2 state.}


\section{Roll cell dynamics}
\label{sect:dynamics-data}

The multiple states in plane Couette flow concern with the dynamics of the roll cells. 
In this section, we will extract from the NS equation two equations that govern the the dynamics of the roll cells.

\subsection{Theoretical derivation}
\label{sect:driv1}

Turbulence has many degrees of freedom, which gives rise to a high dimensional phase space.
Directly studying turbulent dynamics in a high dimensional phase space is difficult (if not impossible).
In order to make progress, we must simplify the NS equation according to the specific flow under consideration.

The NS equation {for RPCF} reads
\begin{equation}
    \frac{D u_i}{D t}=-\frac{1}{\rho}\frac{\partial p}{\partial x_i}+\nu \frac{\partial^2 u_i}{\partial x_k^2}-\epsilon_{i2k}~Ro~u_k,
\end{equation}
where $u_i$ is the fluid velocity in the $i$th Cartesian direction, $D/Dt$ is the total derivative, $i=$1, 2, 3 denote the the streamwise, spanwise, and wall-normal directions.
We use $x$, $y$, $z$ interchangeably with $x_i$; $u$, $v$, $w$ interchangeably with $u_i$; and $\partial_i$ interchangeably with $\partial /\partial x_i$.
%We consider a spanwise rotating Couette flow.
Normalization is by the half of the wall velocity difference $U_w$, and the half channel height $h$.
It follows that $\nu=1/Re_w$, where $Re_w=U_wh/\nu$ is the Reynolds number.
In the main text, we present derivations for plane Couette flow.
Derivations for Taylor Couette flow (which concerns with surface curvature) are presented in appendix \ref{app:TCF}.

%{\bf [STEP 1]} 
First, we filter the NS equation in the streamwise direction at a scale $\sim O(h)$.
%The bold text is for reference purpose.
The filtered streamwise velocity equation reads
\begin{equation}
    \frac{D\tilde{u}_1}{Dt}=-\frac{1}{\rho}\partial_1 \tilde{p}-\partial_j \tilde{T}_{1j} +\nu\partial_j\partial_j \tilde{u}_1-Ro \tilde{u}_3,
    \label{eq:u1-0}
\end{equation}
where $\tilde{\cdot}$ denotes streamwise filtration, and $\tilde{T}_{1j}$ is the turbulent stress.
Define the filtered vorticity $\tilde{\omega}_i=\epsilon_{ijk}\partial_j \tilde{u}_k$.
The streamwise vorticity equation reads
\begin{equation}
    \frac{D\tilde{\omega}_1}{Dt}=\tilde{\omega}_j\partial_j \tilde{u}_1-\epsilon_{1qi}\partial_q\partial_j\tilde{T}_{ij}+\nu\partial_j\partial_j\tilde{\omega}_1+Ro\partial_2\tilde{u}_1.
    \label{eq:w1-0}
\end{equation}
For a spanwise rotating plane Couette flow that features multiple statistically stable states, the flow is approximately streamwise homogeneous.
%This is particularly the case after filtering in the streamwise direction at a scale $\sim O(h)$. 
To focus on the behavior of the large-scale roll cells, we neglect the streamwise derivative and invoke the eddy viscosity model.
{It follows that the vortex stretching term $\tilde{\omega}_j\partial_j \tilde{u}_1$ is zero upon substitution of the vorticity, $\tilde{\omega}_j=\epsilon_{jpq}\partial_p \tilde{u}_q$.}
Hence, \eqref{eq:u1-0} and \eqref{eq:w1-0} lead to
\begin{equation}
    \partial_t\tilde{u}_1+\tilde{u}_j\partial_j\tilde{u}_1=\partial_j\left[(\nu+\nu_t)\partial_j\tilde{u}_1\right]-Ro \tilde{u}_3,
    \label{eq:u1-1}
\end{equation}
and
\begin{equation}
    \partial_t\tilde{\omega}_1+\tilde{u}_j\partial_j\tilde{\omega}_1=\partial_j\left[(\nu+\nu_t)\partial_j\tilde{\omega}_1\right]+Ro\partial_2\tilde{u}_1,
    \label{eq:w1-1}
\end{equation}
where {$j=2,3$ with Einstein summation}, and $\partial_{jj}=\partial_{22}+\partial_{33}$.
%For brevity, $\tilde{\cdot}$ is dropped in the following. 

%{\bf [STEP 2]}
Second, we define a streamfunction $\psi$ such that
\begin{equation}
\begin{split}
\tilde{u}_2&=\partial_3\psi, \\
\tilde{u}_3&=-\partial_2\psi.
\end{split}
\label{eq:psi}
\end{equation}
It follows that $\tilde{\omega}_1=-\partial_j\partial_j\psi$.
%The use of the streamfunction eliminates the pressure.
Plugging \eqref{eq:psi} into \eqref{eq:u1-1} and \eqref{eq:w1-1} leads to
\begin{equation}
    \partial_t\tilde{u}_1
    +\partial_2\tilde{u}_1\partial_3\psi-\partial_3\tilde{u}_1\partial_2\psi
    =\partial_j\left[\frac{1}{R}\partial_j\tilde{u}_1\right]+Ro\partial_2\psi
    \label{eq:u1-2}
\end{equation}
and
\begin{equation}
    \partial_t\partial_j\partial_j\psi
    +\partial_2\partial_j\partial_j\psi\partial_3\psi-\partial_3\partial_j\partial_j\psi\partial_2\psi=\partial_j\left[\frac{1}{R}\partial_j\partial_q\partial_q \psi\right]-Ro\partial_2\tilde{u}_1.
    \label{eq:w1-2}
\end{equation}
In the above equation, we have invoked the effective Reynolds number $R=1/(\nu+\nu_t)$ following, e.g., \cite{anderson2019non}.
As the eddy viscosity is usually not a constant, invoking the effective Reynolds number is to resort to an approximation that works.
Nonetheless, in this context, because the eddy viscosity in the bulk region of a spanwise rotating channel is in fact close to a constant \citep{yang2020mean}, invoking the effective Reynolds number here has a somewhat sounder physical basis.

%{\bf [STEP 3]} 
Third, we project \eqref{eq:u1-2} and \eqref{eq:w1-2} on a statistically stable state
\begin{equation}
\begin{split}
    \psi(x_2,x_3,t)&=a(t)\alpha(x_2,x_3),\\
    \tilde{u}_1(x_2,x_3,t)&=b(t)\beta(x_2,x_3)+U(x_3),
\end{split}
\label{eq:mode}
\end{equation}
where $\alpha$ and $\beta$ represent roll cells of a given wavenumber, and $U(x_3)$ is the mean flow. 
Here, we assume that the mean flow does not depend on a particular state.
This assumption is consistent with the data and simplifies the analysis that follows.
Relaxing this assumption leads to more involved math but essentially the same conclusions, as shown in Appendix \ref{app:meanflow}. 
Plugging \eqref{eq:mode} into \eqref{eq:u1-2} and \eqref{eq:w1-2} leads to
\begin{equation}
\begin{split}
    &\beta \frac{db}{dt}+ab\partial_2\beta\partial_3\alpha-ab\partial_3\beta \partial_2\alpha-a\partial_3U \partial_2\alpha\\
    =&\frac{1}{R}~b\partial_j\partial_j\beta+\frac{1}{R}\partial_3\partial_3U+Ro~a\partial_2\alpha
\end{split}
\label{eq:u1-3}
\end{equation}
and
\begin{equation}
\begin{split}
    &\partial_j\partial_j\alpha \frac{da}{dt}+a^2(\partial_2\partial_j\partial_j\alpha\partial_3\alpha-\partial_3\partial_j\partial_j\alpha\partial_2\alpha)\\
    =&\frac{1}{R}~a\left[\partial_2^4\alpha+2\partial_2^2\partial_3^2\alpha+\partial_3^4\alpha\right]-Ro~b\partial_2\beta
\end{split}
\label{eq:w1-3}
\end{equation}
Equations \eqref{eq:u1-3} and \eqref{eq:w1-3} govern the dynamics of roll cells.
%and map the flow dynamics on $a$, $b$.
In \eqref{eq:u1-3} and \eqref{eq:w1-3}, we have left the $\alpha$ and $\beta$ as generic functions of $y$ and $z$, and, importantly, we have retained non-linearity.

%In arriving at \eqref{eq:u1-3} and \eqref{eq:w1-3}, we have assumed first that the streamwise filtered streamwise velocity $u$ and streamwise vorticity $\omega_1$ are streamwise homogeneous, and second that the mean velocity profile does not depend on the states.

\begin{figure}
\centering
\vspace{2mm}
\includegraphics[width=0.8\textwidth]{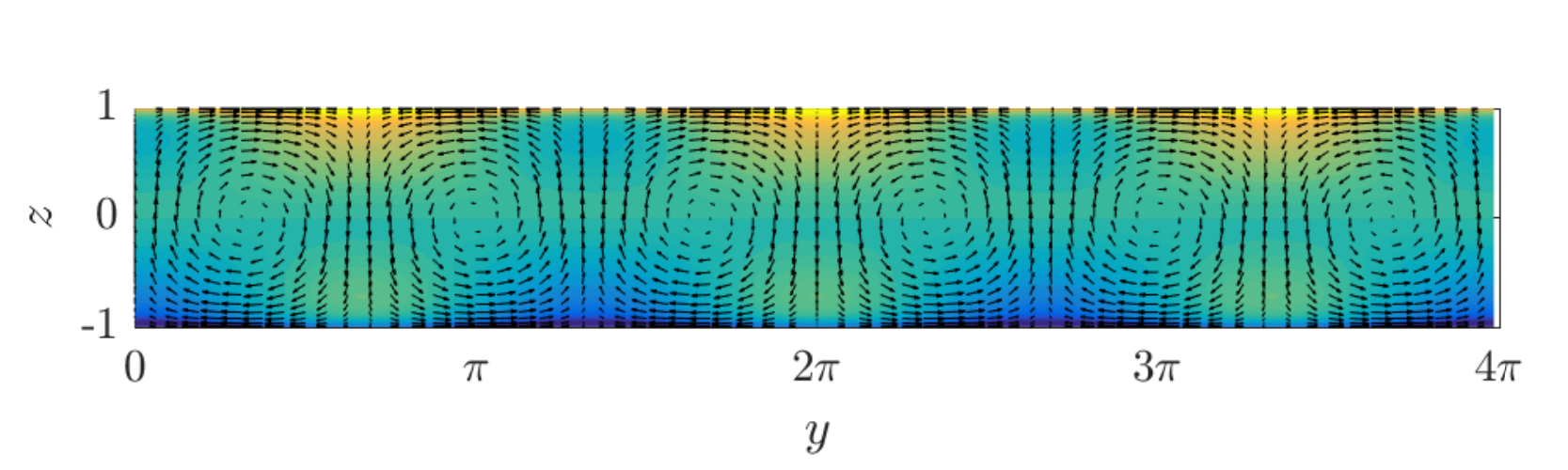}
\caption{Equation \eqref{eq:ansatz-ab}. Contours are for $u$ and vectors are for $v$ and $w$.
Again, the exact contour values are not important.
}
\label{fig:ansatz}
\end{figure}

\begin{figure}
\centering
\vspace{2mm}
\includegraphics[width=0.4\textwidth]{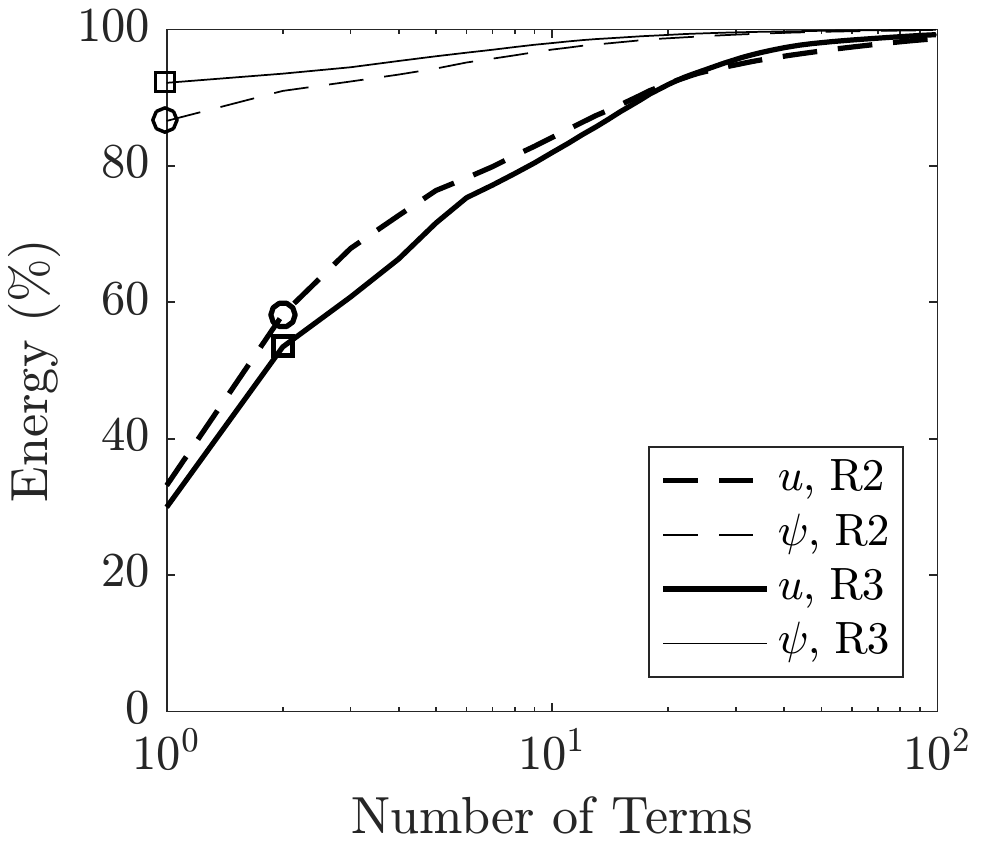}
\caption{The energy contained in the ansatz as a function of $n$, where $n$ is the number of terms in the ansatz. 
The thick lines are for $u$.
The thin lines are for $\psi$.
The dashed lines are for R2.
The solid lines are for R3.
}
\label{fig:energy}
\end{figure}

%{\bf[STEP 4]} 
Fourth, we need ansatzes for $\alpha$ and $\beta$ in order to advance.
The DNS solution is by itself a good ansatz, but the use of DNS data necessarily involves numerical tools in the following derivations, which will be postponed to Appendix \ref{app:data-ansatz}.
Here, we use the following analytical ansatz to approximate the multiple states  
\begin{equation}
\begin{split}
    \alpha&=-\sin\left(\frac{\pi}{2}ky\right)\cos\left(\frac{\pi}{2}z\right),\\ ~~~\beta&=-\cos\left(\frac{\pi}{2}ky\right)\left[1-\cos\left(\pi z\right)\right]. 
\end{split}
\label{eq:ansatz-ab}
\end{equation}
Figure \ref{fig:ansatz} shows the ansatz for $k=3/\pi$, which corresponds to R3.
As we can see, the ansatz compares well with the data, i.e., figure \ref{fig:xia-R23} (b).
Quantitatively, our analytical ansatz contains about 55\% of the energy in $\tilde{u}$ and about 90\% of the energy in $\tilde{v}$ and $\tilde{w}$.
The next most energetic Fourier mode contains about 7\% of the energy in $\tilde{u}$ and 1\% of the energy in $\tilde{v}$ and $\tilde{w}$.
For R2, our analytical ansatz contains about 60\% of the energy in $\tilde{u}$ and about 85\% of the energy in $\tilde{v}$ and $\tilde{w}$.
The next most energetic Fourier mode contains about 7\% of the energy in $\tilde{u}$ and about 4\% of the energy in $\tilde{v}$ and $\tilde{w}$.
One could obtain more realistic ansatzes by including additional terms in \eqref{eq:ansatz-ab}.
Figure \ref{fig:energy} shows the energy contained in the ansatzes as we include more Fourier modes.
As one would expect, accuracy comes at the price of brevity.
In appendix \ref{app:data-ansatz}, we resort to DNS for an ansatz, and we will see that our conclusions are not affected.

\subsection{Dynamics of the stable states}
\label{sect:driv2}

\begin{figure}
\centering
\vspace{2mm}
\includegraphics[width=0.4\textwidth]{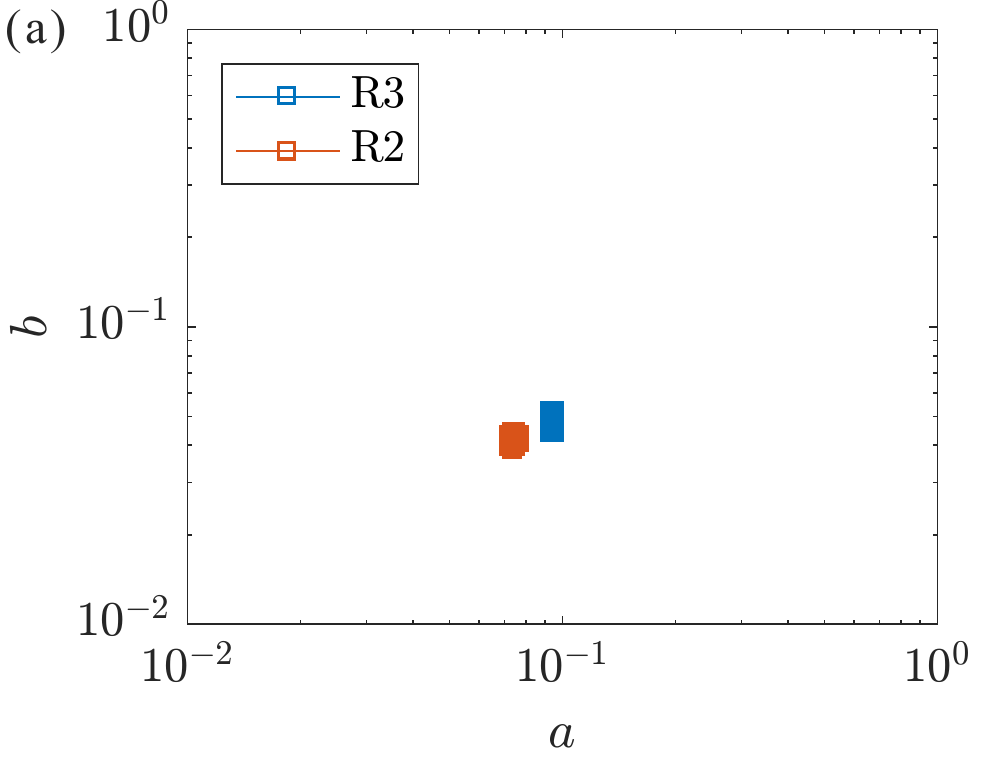}~~\includegraphics[width=0.4\textwidth]{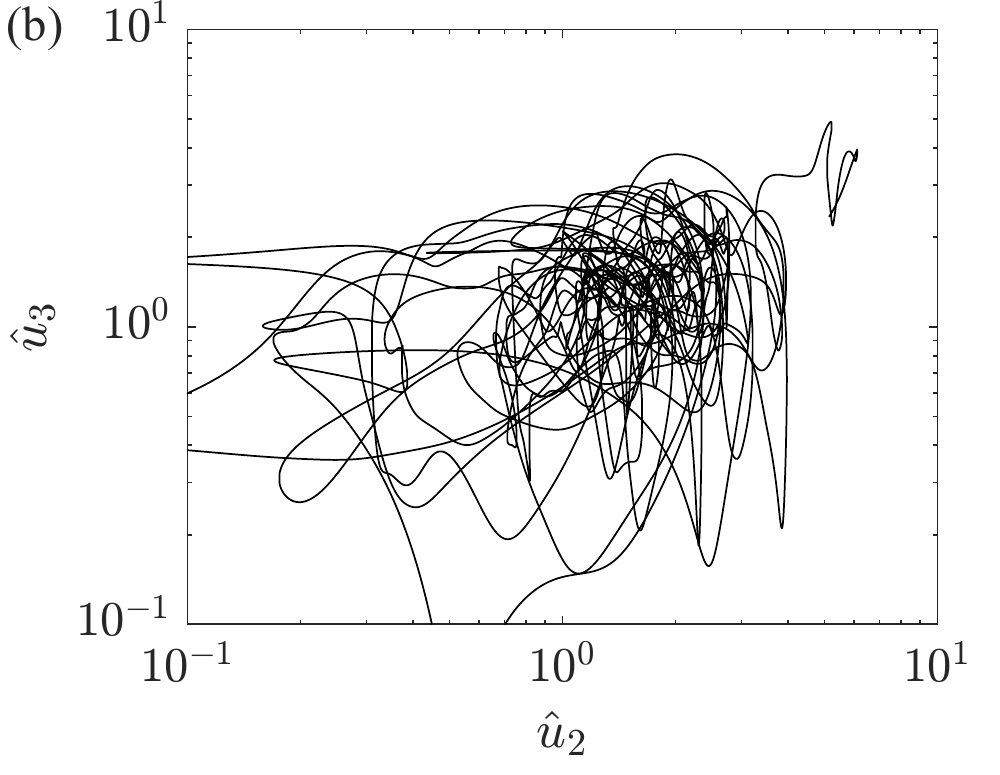}
\caption{(a) Trajectories of R2 and R3 in the phase space of $a$ and $b$.
(b) Trajectory of isotropic turbulence in the sub-phase space of two Fourier modes $\hat{u}_2$ and $\hat{u}_3$.
%with $k=2$ and $k=3$.
The $x$ and $y$ axes in both figures span two decades.
}
\label{fig:phase-comp}
\end{figure}
Before we proceed with our derivation, we take a closer look at the flow dynamics near the two stable states in R2 and R3 {so that we have in mind what to expect}.
Equations \eqref{eq:u1-3} and \eqref{eq:w1-3} map the roll cell dynamics to two modes: $a$ and $b$.
Projecting the DNS data on the ansatzes in \eqref{eq:ansatz-ab}, in figures \ref{fig:phase-comp} (a), we plot the trajectories of R2 and R3 in the phase space of $a$ and $b$.
For comparison, in figure \ref{fig:phase-comp} (b), we plot the trajectory of an isotropic flow in the sub-phase-space of two Fourier modes.
The integral length of this isotropic flow is $L$.
The two Fourier modes correspond to the length scales of $L/2$ and $L/8$ (relatively large scale modes).
Details of the data in figure \ref{fig:phase-comp} (b) are presented in Appendix \ref{app:SM}.

Comparing figures \ref{fig:phase-comp} (a, b), the trajectory in figure \ref{fig:phase-comp} (b) visits a large area in the phase space and is easily ergodic, but the two trajectories in figure \ref{fig:phase-comp} (a) are confined in two small regions, and the two regions do not overlap, thereby leading to two independent statistically stable states. 
%This is the physics we hope to capture.
%The presence of multiple states, or non-ergodicity, is because the flow does not have enough energy to get out of a statistically stable state. 
%As we will see in later section, our theory reflects the above understanding.

\begin{figure}
\centering
\includegraphics[width=0.3\textwidth]{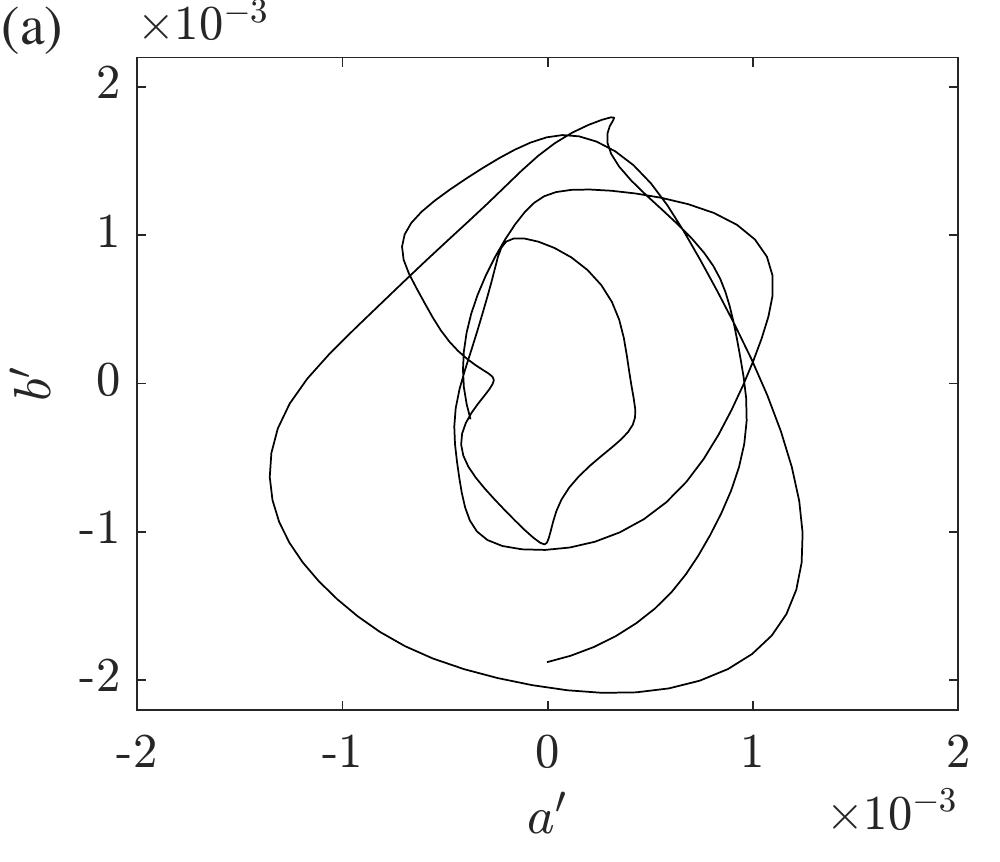}~~\includegraphics[width=0.3\textwidth]{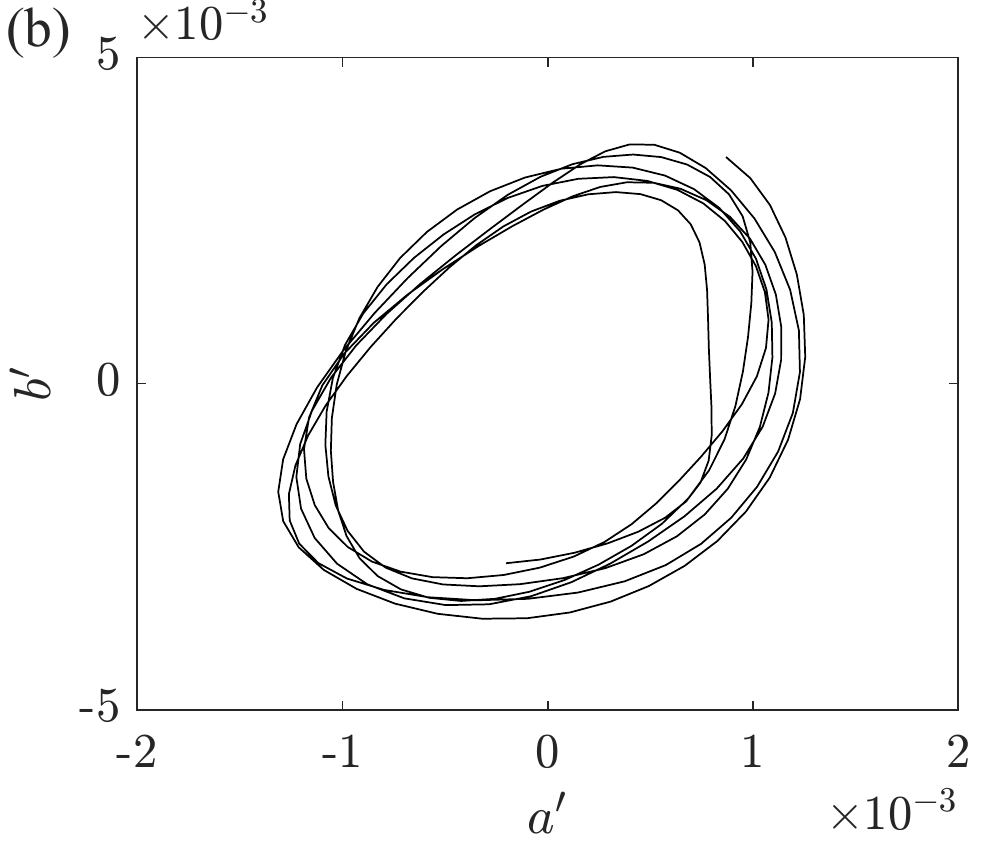}
\caption{Zoomed view of the trajectories of (a) R2 and (b) R3 in the phase space of $a^\prime$ and $b^\prime$.
}
\label{fig:phase-near}
\end{figure}

\begin{figure} 
\centering
\includegraphics[height=1.2in]{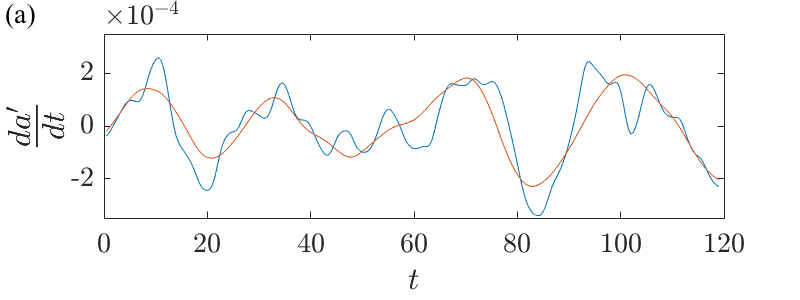}\\
\includegraphics[height=1.2in]{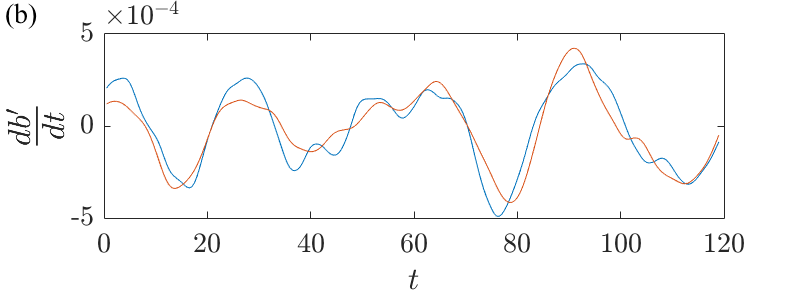}
\caption{The time evolution of (a) $a^\prime$ and (b) $b^\prime$ in R2 and fits to linear dynamics.
}
\label{fig:ab2}
\end{figure}

\begin{figure} 
\centering
~~~~~~~~~~~~~~~~~~~~~~\includegraphics[height=1.2in]{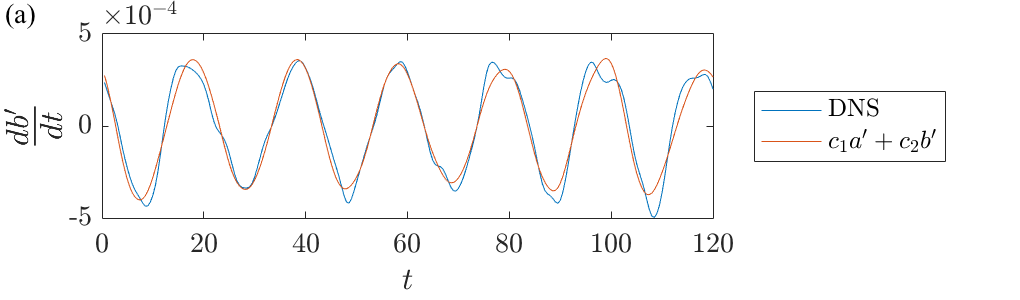}
\includegraphics[height=1.2in]{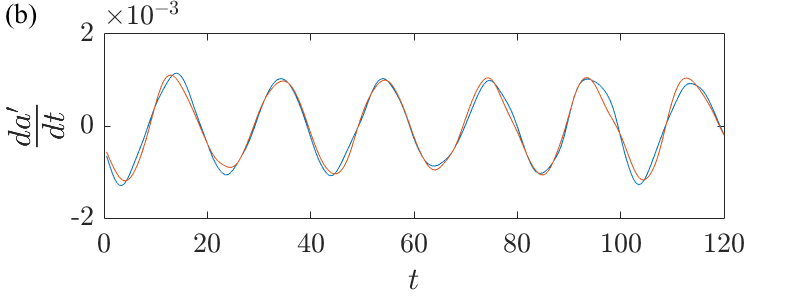}
\caption{The time evolution of (a) $a^\prime$ and (b) $b^\prime$ in R3 and fits to linear dynamics.
}
\label{fig:ab3}
\end{figure}

%We study the roll cell dynamics near the two statistically stable states.
Figures \ref{fig:phase-near} (a, b) show a zoom-in view of the trajectories of R2 and R3 respectively in the phase space of $a^\prime$ and $b^\prime$, where the superscript $\prime$ denotes temporal fluctuations.
%In figure \ref{fig:phase-near}, we have removed the large-scale drift in the signal.
Limit-cycle like trajectories are found in both R2 and R3.
For reasons that will be clear in the next subsection, we fit $da^\prime/dt$ and $db^\prime/dt$ to the linear dynamics, i.e., $c_1a^\prime+c_2b^\prime$, where $c_1$, $c_2$ are constants.
In figures \ref{fig:ab2} and \ref{fig:ab3}, the data is compared to the linear dynamics, and the two are remarkably similar, suggesting that the flow near the two stable states is govern by linear dynamics. 
Ergodicity relies on sufficiently large fluctuations, and in a turbulent flow system, large fluctuations are often results of non-linear interactions.
The fact that the evolution of $a^\prime$ and $b^\prime$ follows linear dynamics explains, from a phenomenological standpoint, why the flow is trapped within a stable state.
The challenge remains that whether we could theoretical show that $a$ and $b$ follow linear dynamics from the NS equation.

%We have moved more detailed analysis of the roll cell dynamics in the vicinity of the statistically stable states to Appendix \ref{app:dynamics}, as it is not essential to the discussion in the main text.

\subsection{Truncated Galerkin projection and bifurcation analysis}
\label{sect:TGP}

We will proceed with our derivation.
In section \ref{sect:driv1}, we were at step four.
In section \ref{sect:driv2}, we see that the data suggests linear dynamics near a stable state.
In this section, we will see if our derivation gives rise to the observed linear dynamics.
The fifth step is to plug \eqref{eq:ansatz-ab} into \eqref{eq:u1-3} and \eqref{eq:w1-3} and project the two equations on to the ansatzes in \eqref{eq:ansatz-ab}.
Here, projecting a function $f$ on $\alpha$ is to integrate 
\begin{equation}
\int_0^{L_y}\int_{-1}^{1} f\cdot \alpha dz dy.
\end{equation}
After some (long) algebra, we arrive at
\begin{equation}
    \frac{db}{dt}=\left[\frac{4k}{9}Ro+\frac{k\pi}{3}\int_{-1}^1\sin^2\left(\frac{\pi}{2}z\right)\cos\left(\frac{\pi}{2}z\right)\frac{dU}{dz} dz\right]a
    -\left(k^2+\frac{4}{3}\right)\frac{1}{R}\left(\frac{\pi}{2}\right)^2b
    \label{eq:u1-4}
\end{equation}
and
\begin{equation}
    \frac{da}{dt}=-\frac{1}{R}(k^2+1)(\pi/2)^2a
    -\frac{4}{3}\frac{k}{k^2+1}Ro(\pi/2)^{-2}b,
    \label{eq:w1-4}
\end{equation}
{which is linear dynamics.}
In arriving at the above two equations, we invoke $\sin(\pi/2~kL_y)=0$ since there can only be integer pairs of roll cells in the domain.
It is worth noting that truncated Galerkin projection does not preclude nonlinearity.
For instance, \cite{anderson2019non} obtained a truncated system with nonlinear terms. 
%A lack of nonlinearity here has more profound physical implications, as discussed in Appendix \ref{app:dynamics}.
The integral in \eqref{eq:u1-4} is
\begin{equation}
\int_{-1}^1\sin^2\left(\frac{\pi}{2}z\right)\cos\left(\frac{\pi}{2}z\right)\frac{dU}{dz} dz=-\int_{-1}^1 U~\frac{\pi}{2} \sin\left(\frac{\pi}{2}z\right)\left[2-3\sin^2\left(\frac{\pi}{2}z\right)\right]dz \approx 0.27
\label{eq:integral}
\end{equation}
for both R2 and R3.
As is clear from \eqref{eq:integral}, the integral concerns with $U$ not its derivative.
%Equations \eqref{eq:u1-4} and \eqref{eq:w1-4} govern the roll cells.
%The dynamical system truncates any interaction among the roll cells and the mean flow.
%In appendix \ref{app:B}, we include additional terms and model the dynamics of roll cells as well as the mean flow.
%We will see that it does not affect our conclusion

\begin{figure}
\centering
\vspace{2mm}
\includegraphics[width=0.65\textwidth]{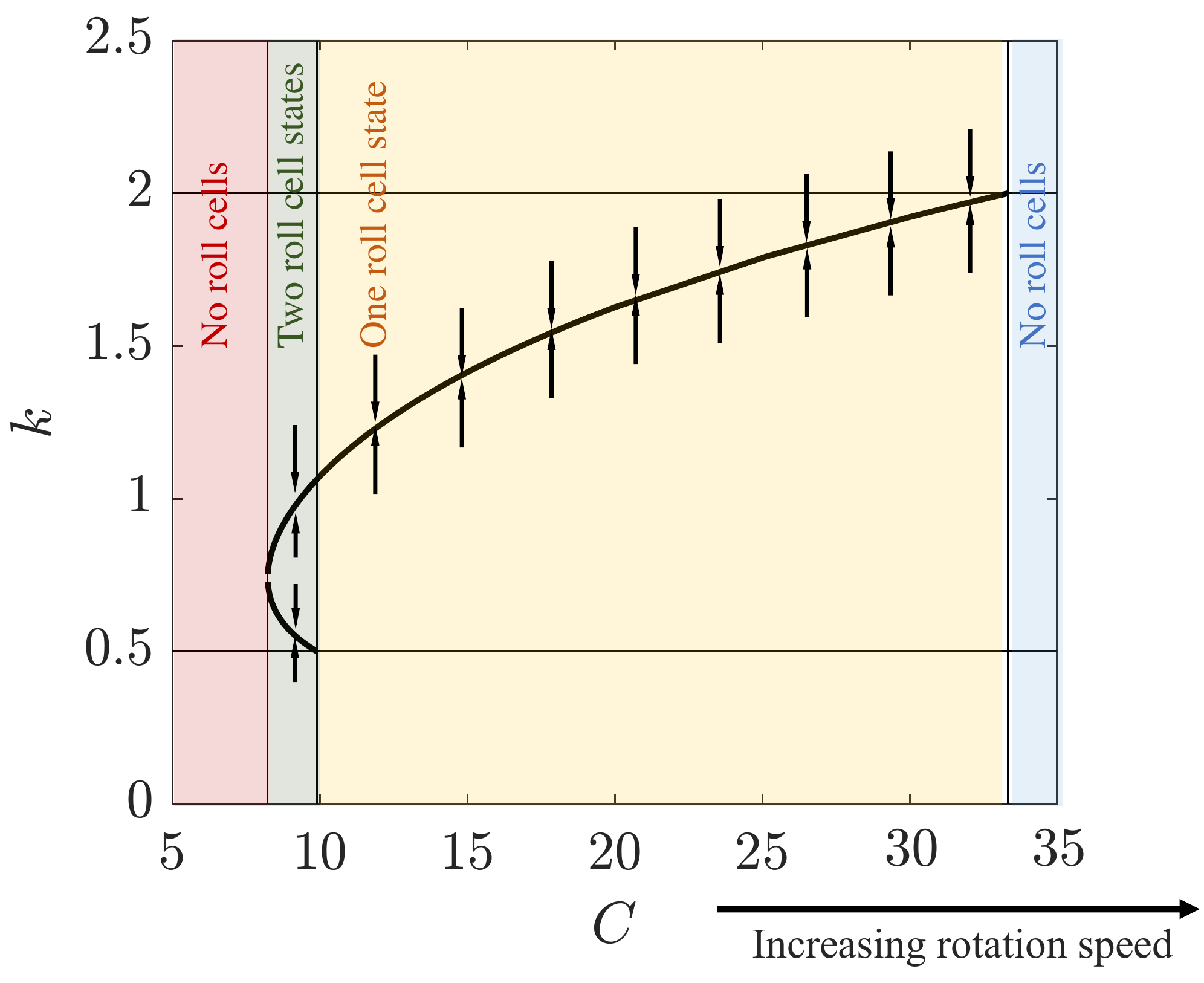}
\caption{Bifurcation diagram. 
Per \eqref{eq:lam}, $C\approx -\left(0.039 Ro^2+0.025 Ro\right)R^2$.
{For the condition considered here, i.e., near $Ro=-0.2$, $C$ is an increasing function of the rotation speed $\left|Ro\right|$.}
}
\label{fig:bi2}
\end{figure}

%{\bf [STEP 6] Bifurcation analysis} 
We are now at the last step.
Sixth, we conduct bifurcation analysis of the ODEs in \eqref{eq:u1-4} and \eqref{eq:w1-4}.
If multiple statistically stable state exist, $da/dt=0$ and $db/dt=0$ must have multiple non-trivial solutions.
Hence, if multiple states exist, the eigen values of the ODEs in \eqref{eq:u1-4} and \eqref{eq:w1-4} must be such that
\begin{equation}
\lambda_{1,2}=0,
\end{equation}
for non-trivial $a$ and $b$.
This necessarily leads to:
\begin{equation}
    (k^2+1)^2(k^2+4/3)-Ck^2=0,~~~C\approx- \left(0.039 Ro^2+0.025 Ro\right)R^2
    \label{eq:lam}
\end{equation}
Equation \eqref{eq:lam} is a cubic equation of $k^2$ (note that $Ro<0$ and $C>0$).
For certain $C$, i.e., for certain flow condition, there are two physically viable roots:
\begin{equation}
\begin{split}
    &k_1^2=h^{1/3} + gh^{-1/3} - 10/9,\\
    &k_2^2=-i\sqrt{3/4}\times( h^{1/3}- gh^{-1/3}) -1/2\times(h^{1/3} + gh^{-1/3})-10/9,
\end{split}   
\label{eq:roots}
\end{equation}
where $f=5C/9+1/729$, $g=C/3+1/81$, $h=(f^2 - g^3)^{1/2} - f$, $i=\sqrt{-1}$, $i^{1/3}=\exp(i\cdot \pi/6)$, $(-i)^{1/3}=\exp(-i\cdot \pi/6)$.
Here, $k_2^2$ is real and is positive despite the appearance of the unit imaginary number $i$. 
Hydrodynamic stability does not allow for very slim nor very fat roll cells \citep{drazin2004hydrodynamic}.
Determining the limits of roll cells' aspect ratio falls out of the scope of this paper.
Here, we arbitrarily cutoff at $k=0.5$ and $k=2$, constraining that the roll cells' heights to be no more than twice of their widths and no less than half of their widths.
Figure \ref{fig:bi2} shows the bifurcation diagram, where we plot $k_1$ and $k_2$ as a function of $C$.
We concludes our derivation here. 

Before we proceed with predictions, we make two observations.
First, truncated Galerkin projection is known to preserve non-linearity in the equation, however, the application of truncated Galerkin projection to this particular problem has led to linear dynamics.
Second, \eqref{eq:lam} is, strictly speaking, a sextic polynomial, which, in principle, has 6 roots.
Nonetheless, it turns out that if one regards $k^2$ as the unknown, the equation reduces to a cubic one; and it also happens that one of the three roots of the cubic equation is always negative, leaving us only two physically viable roots.

\section{Predictions of the theory}
\label{sect:results}

\subsection{multiple states in an infinitely large domain}

\begin{figure}
\centering
\vspace{2mm}
\includegraphics[width=0.4\textwidth]{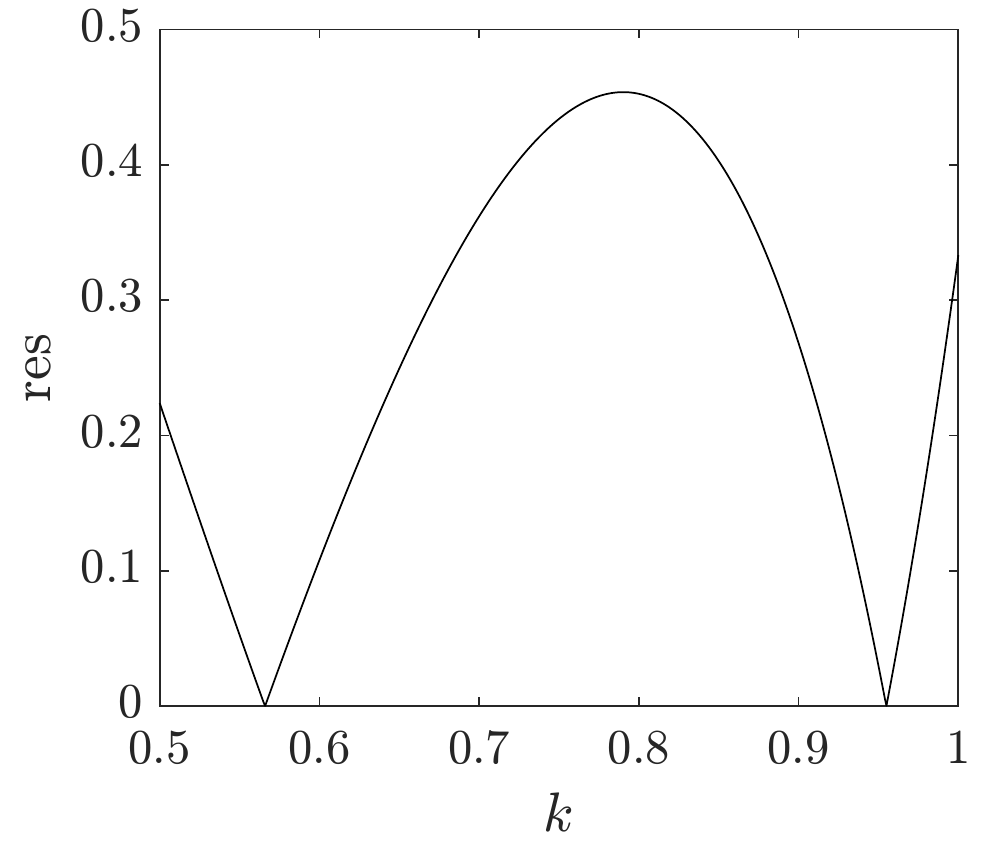}
\caption{Residual of \eqref{eq:lam} for $C=9$.
}
\label{fig:residual}
\end{figure}

%An infinitely large domain admits all wavenumber.
Per figure \ref{fig:bi2}, for small values of $Ro$, the flow is in the red zone and there are no physically viable root to Eq. \eqref{eq:lam}.
It follows that, for a non-rotating plane Couette flow, multiple states that feature roll cells with different wavenumbers are not possible.
For a flow that is in the green zone, i.e., for a moderate rotation number, \eqref{eq:lam} has two roots between 0.5 and 2, leading to multiple states.
In the yellow zone, i.e., at high rotation speeds, \eqref{eq:lam} has one root between 0.5 and 2.
This means that while roll cells could still be found, they could only be found at one wavenumber.
In the blue zone, i.e., at very high rotation speeds, \eqref{eq:lam} does not have a root between 0.5 and 2, and no roll cells could be found.
For a $C$ that gives rise to multiple states, e.g., $C=9$, we plot the residual of the cubic equation \eqref{eq:lam} in figure \ref{fig:residual}.
The two statistically stable states yields 0.
The residual measures the distance of a state from the equilibrium and therefore can be interpreted as a measure of the energy level.
From figure \ref{fig:residual}, we see that the large energy barrier between the stable states locks the flow at one state, leading to non-ergodicity. 
The system would have been ergodic if the energy barrier is low or if the flow has enough energy to go from one state to the other.

\subsection{multiple states in a finite domain}

\begin{figure}
\centering
\vspace{2mm}
\includegraphics[width=0.33\textwidth]{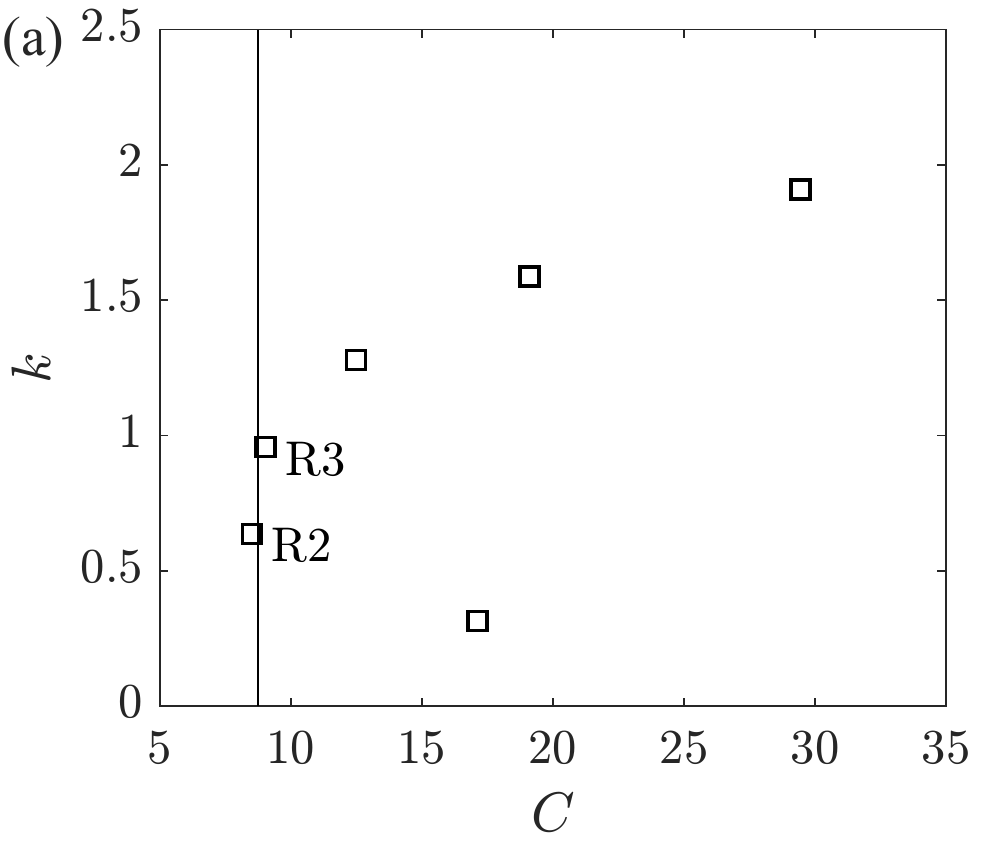}~\includegraphics[width=0.33\textwidth]{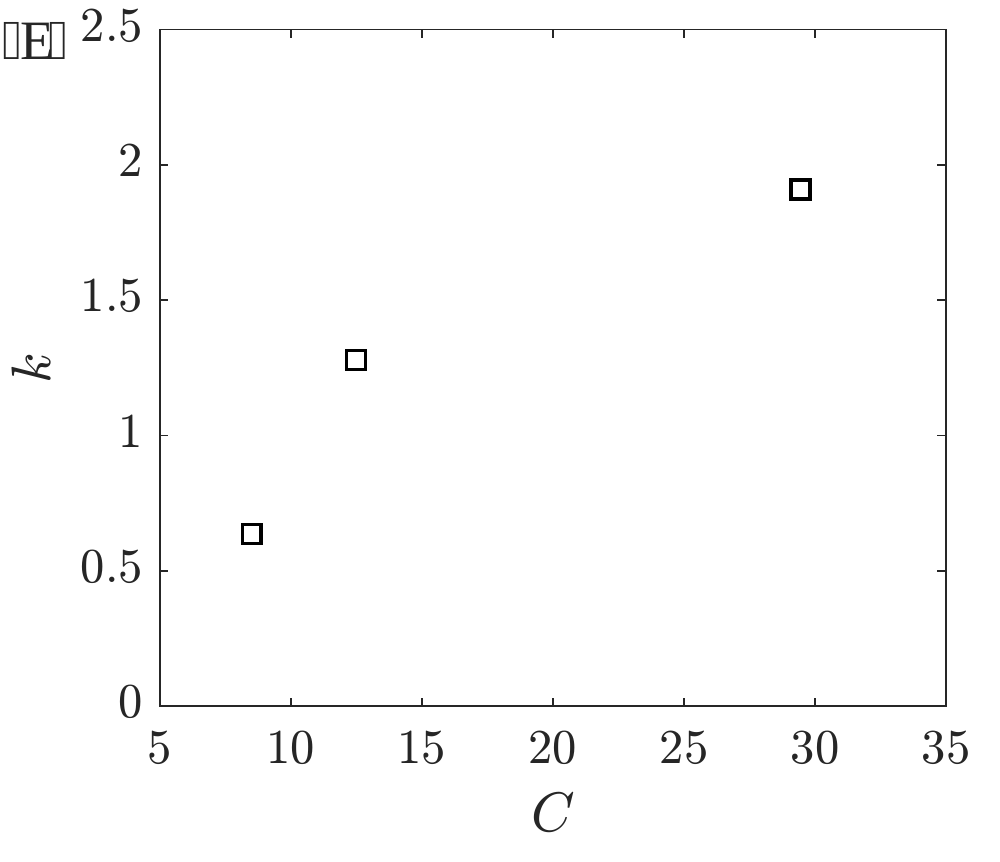}~\includegraphics[width=0.33\textwidth]{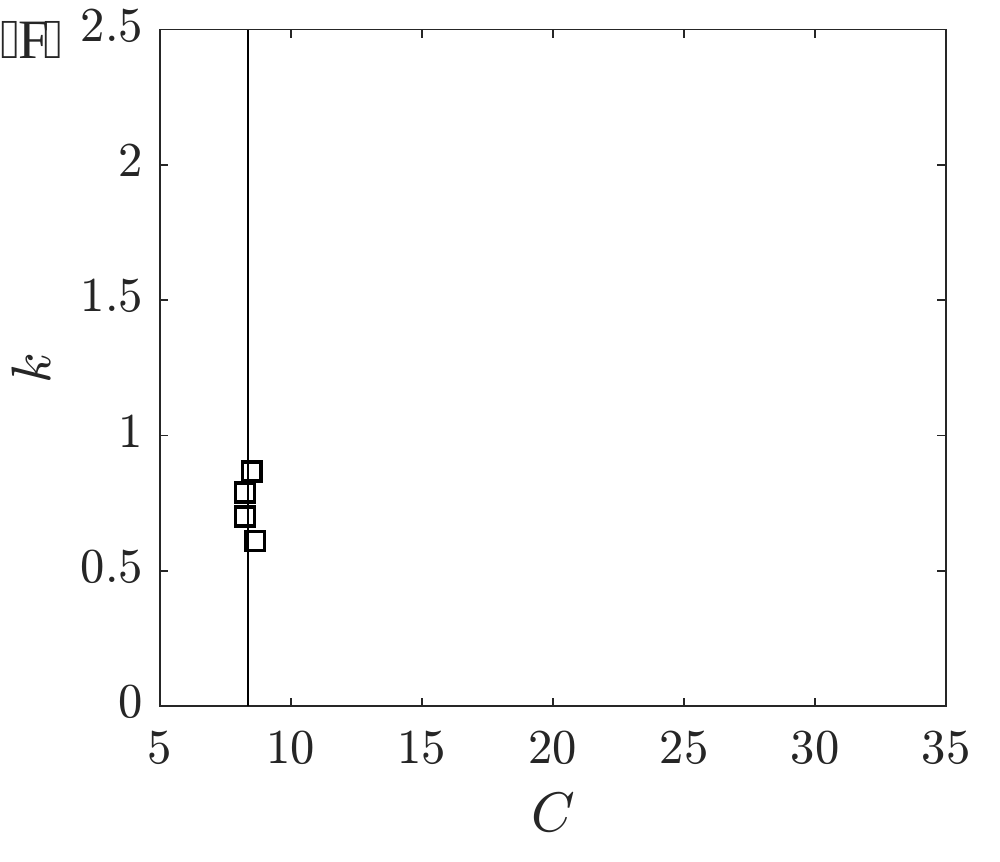}
\caption{(a) Bifurcation diagram for a $L_y=4\pi$ domain.
The wavenumber only takes value at $k=n/\pi$.
(b) Bifurcation diagram for a $L_y=2\pi$ domain.
The wavenumber only takes value at $2n/\pi$.
(c) Bifurcation diagram for a finite domain. The four wavenumbers are equally distanced from the equilibrium.
}
\label{fig:k3}
\end{figure}

Finite domain admits only a few discrete wavenumbers.
For a domain that extends about $L_y=4\pi$ in the spanwise direction, the wavenumber $k$ can only take vales $1/\pi$, $2/\pi$, $3/\pi$, ..., as sketched in figure \ref{fig:k3} (a).
According to figure \ref{fig:k3} (a),  for a $L_y=4\pi$ domain, $k=2/\pi$ and $k=3/\pi$ are two possible statistically stable states.
Had the domain size been $L_y=2\pi$, no multiple states would be possible, as shown figure \ref{fig:k3} (b). 
This explains the selection of certain state by domain.
Last, observe the slight imbalance between $k=2/\pi$ and $k=3/\pi$.
This imbalance makes R2 the preferred state if one increases the rotation speed from 0 to Ro=0.2 (i.e., $C$ increases from 0) and R3 the preferred state if one decreases the rotation speed from 0.5 to 0.2 (i.e., $C$ decreases from a large number), thereby explaining the selection of one state over the other by the initial condition \citep{huang2019hysteresis}.
Allowing for some imbalance, the bifurcation diagram would in fact allow for three and four multiple states in a finite domain, as sketched in figure \ref{fig:k3} (c). 

\section{Conclusions}
\label{sect:conclusion}

We present an analytically tractable derivation that gives rise to multiple states in rotating flows.
%While multiple states have led many interesting discussion, previous works of multiple states were by and large phenomenological.
%This work applies Galerkin Truncation 
We show that non-ergodicity in a spanwise rotating channel is because the flow does not have the energy to escape a statistically stable state. 
%In Appendix \ref{app:dynamics}, we show that the lack of energy is because of the linear roll cell dynamics near the statistically stable states. 
According to this derivation,  spanwise rotating plane Couette flow in an infinitely large domain can have two and only two statistically stable states.
Finite domains (albeit with periodic boundary conditions in the horizontal directions) admit wavenumbers at a few discrete values.
This gives rise to the observed selection of certain statistically stable state by differently sized domain and different initial conditions.
A few straight-forward extensions of our theory are presented in the appendices.
They either involve the use of numerical tools, which enable us to handle more degrees of freedom, or long math derivations.
%Nonetheless, the discussion in the appendices shows the robustness of our conclusions.
Refining this derivation to a point where it could make more quantitative predictions, like the exact rotation number from which we can observe multiple states, is tantalizingly out of reach at this moment and will be the focus of the future research.

\vspace{3mm}
\section*{Acknowledgement}
Yang hopes to thank Charle Meneveau, John Cimbala and Rob Kunz for useful inputs. 
Yang acknowledges ONR for financial support. Xia acknowledges the support from NSFC (Grant Nos. 11822208 and 11772297)

\section*{Appendix}
\appendix
\section{Taylor-Couette flow}
\label{app:TCF}

\begin{figure}
\centering
\vspace{2mm}
\includegraphics[width=0.85\textwidth]{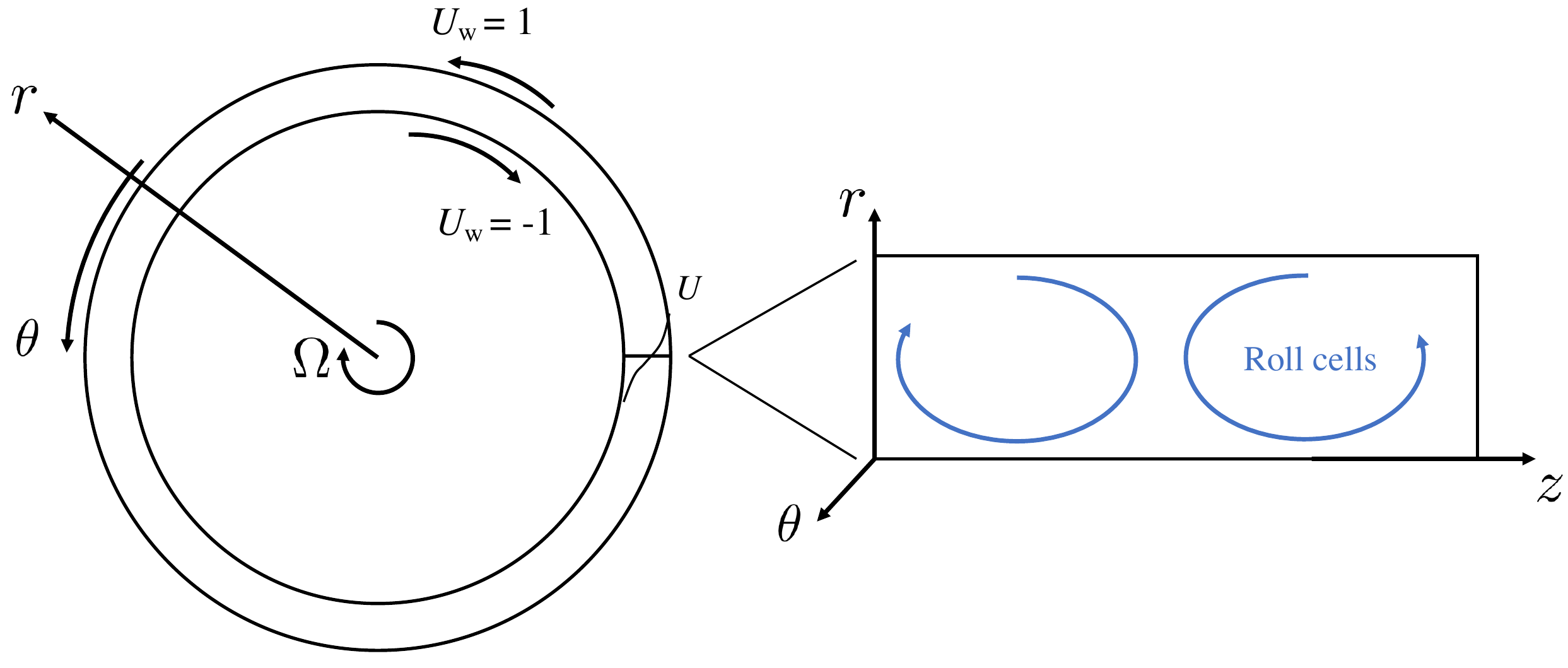}
\caption{A sketch of Taylor-Couette flow.
}
\label{fig:TC}
\end{figure}

Figure \ref{fig:TC} is a sketch of Taylor Couette flow. 
$\theta$ is the ``streamwise'' direction, $z$ is the ``spanwise'' direction, and $r$ is the  ``wall-normal'' direction.
Without loss of generality, the radius of the outer cylinder is $r=r_0+1$ and the radius of the inner cylinder is $r=r_0-1$.
The azimuthal filtered flow is approximately homogeneous in the $\theta$ direction.
Repeat the first step but for Taylor-Couette flow, we get two equations that governs the dynamics of $\tilde{u}_\theta$ and  $\tilde{\omega}_\theta$:
\begin{equation}
\begin{split}
    \frac{\partial \tilde{u}_\theta}{\partial t}+\tilde{u}_r\frac{\partial \tilde{u}_\theta}{\partial r}+\frac{\tilde{u}_r\tilde{u}_\theta}{r}+\tilde{u}_z\frac{\partial \tilde{u}_\theta}{\partial z}&=\frac{1}{R}\left[\frac{1}{r}\frac{\partial }{\partial r}\left(r\frac{\partial \tilde{u}_\theta}{\partial r}\right)-\frac{\tilde{u}_\theta}{r^2}+\frac{\partial^2 \tilde{u}_\theta}{\partial z^2}\right]-Ro \tilde{u}_r\\
    \frac{\partial \tilde{\omega}_\theta}{\partial t}+\tilde{u}_r\frac{\partial  \tilde{\omega}_\theta}{\partial r}+\tilde{u}_z\frac{\partial  \tilde{\omega}_\theta}{\partial z}&=\frac{1}{R}\left[\frac{\partial^2  \tilde{\omega}_\theta }{\partial r^2}+\frac{1}{r}\frac{\partial  \tilde{\omega}_\theta}{\partial r}+\frac{\partial^2 \tilde{\omega}_\theta}{\partial z^2}\right]+Ro\frac{\partial \tilde{u}_\theta}{\partial z}.
\end{split}
\label{eq:TC-uomg}
\end{equation}
Observe that for $r_0\gg 1$, \eqref{eq:TC-uomg} reduces to \eqref{eq:u1-1} and \eqref{eq:w1-1} (upon a coordinate transformation that maps $\theta$ to $x$, $z$ to $y$, and $r$ to $z$).
A streamline function $\psi$ could be defined:
\begin{equation}
\begin{split}
    \tilde{u}_r&=-\frac{1}{r}\frac{\partial \psi}{\partial z},\\
    \tilde{u}_z&=\frac{1}{r}\frac{\partial \psi}{\partial r},
\end{split}
\label{eq:TC-vw}
\end{equation}
and it follows that the vorticity in the $\omega$ direction is
\begin{equation}
     \tilde{\omega}_\theta=-\frac{1}{r}\left[\frac{\partial^2 \psi}{\partial z^2}+\frac{\partial^2 \psi}{\partial r^2}-\frac{1}{r}\frac{\partial \psi}{\partial r}\right].
\label{eq:TC-omg}
\end{equation}
Repeating step two and plugging \eqref{eq:TC-vw} and \eqref{eq:TC-omg} into \eqref{eq:TC-uomg} lead to two equations for $\tilde{u}_\theta$ and $\psi$, similar to \eqref{eq:u1-2} and \eqref{eq:w1-2}.

Further advancements will require ansatzes for the roll cells.
Without loss of generality, a possible ansatz is
\begin{equation}
\begin{split}
    \psi&=a(t) \alpha(r,z), ~~~\alpha(r,z)=-r_0 \sin\left(\frac{\pi}{2}kz\right)\cos\left(\frac{\pi}{2}r\right);\\
    \tilde{u}_\theta&=b(t) \beta(r,z), ~~~\beta(r,z)=-\cos\left(\frac{\pi}{2}kz\right)\left(1-\cos\left(\pi r\right)\right).\\
\end{split}
\label{eq:TC-ab}
\end{equation}
Proceeding with steps five will lead to two ODEs for $a(t)$ and $b(t)$, whose non-trivial nodes give the statistically stable states.
However, such analysis involves integrals that could not be evaluated analytically.
Also, considering that we do not have access to the Taylor-Couette flow data, we will not pursue this analysis in this paper.

\section{Account for interactions between the mean flow and the roll cells}
\label{app:meanflow}

We add an additional term $c(t) \gamma(x_3)$ to the ansatz of $\tilde{u}_1$ in order to model the interactions between the mean flow and the roll cells:
\begin{equation}
\begin{split}
    \psi(x_2,x_3,t)&=a(t)\alpha(x_2,x_3),\\ \tilde{u}_1(x_2,x_3,t)&=b(t)\beta(x_2,x_3)+c(t)\gamma(x_3)+U(x_3).
\end{split}
\label{eq:mode2}
\end{equation}
Repeating step three and plugging \eqref{eq:mode2} into \eqref{eq:u1-2} lead to
\begin{equation}
\begin{split}
    &\beta \frac{db}{dt}+\gamma\frac{dc}{dt}+ab\partial_2\beta\partial_3\alpha-ab\partial_3\beta \partial_2\alpha-a\partial_3U \partial_2\alpha
    -ac\partial_3\gamma \partial_2\alpha\\
    =&\frac{b}{R}\partial_j\partial_j\beta+\frac{1}{R}\partial_3\partial_3U+\frac{c}{R}\partial_3\partial_3\gamma+Ro~a\partial_2\alpha,
\end{split}
\label{eq:u1-32}
\end{equation}
and plugging \eqref{eq:mode2} into \eqref{eq:w1-2} leads to \eqref{eq:w1-3}.
Next, we repeat step four by using the same ansatze in \eqref{eq:ansatz-ab} for $\alpha$ and $\beta$. 
The following ansatz for $\gamma$
\begin{equation}
    \gamma=-\sin(\pi z).
\end{equation}
Repeating step 5 and projecting \eqref{eq:u1-32} onto $\beta$ and $\gamma$ leads to
\begin{equation}
\begin{split}
    \frac{db}{dt}&=
    \left[\frac{4k}{9}Ro+\frac{0.14\pi k}{3}\right]a
    -\left(k^2+\frac{4}{3}\right)\frac{1}{R}\left(\frac{\pi}{2}\right)^2b
    +\frac{4\pi k}{45}ac,\\
    \frac{dc}{dt}&=-\frac{\pi^2}{R}c-\frac{2\pi k}{15}ab.
\end{split}
\label{eq:u1-42}
\end{equation}
Projecting \eqref{eq:w1-3} onto $\alpha$ leads to \eqref{eq:w1-4}.
Equations \eqref{eq:u1-42} and \eqref{eq:w1-4} govern the roll cell dynamics.
The ODEs in \eqref{eq:u1-42} and \eqref{eq:w1-4}, however, have the same non-trivial node as \eqref{eq:u1-4} and \eqref{eq:w1-4} with $c=0$.

\section{Data-based ansatz for roll cells}
\label{app:data-ansatz}

\begin{figure}
\centering
\vspace{2mm}
\includegraphics[height=1.5in]{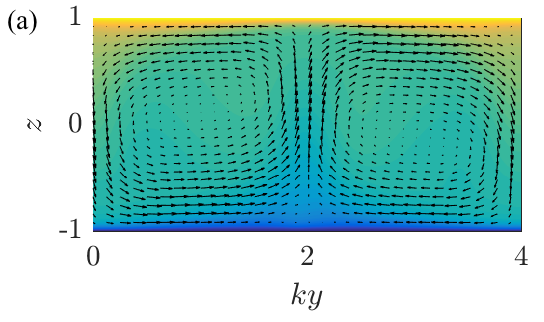}~~~~\includegraphics[height=1.5in]{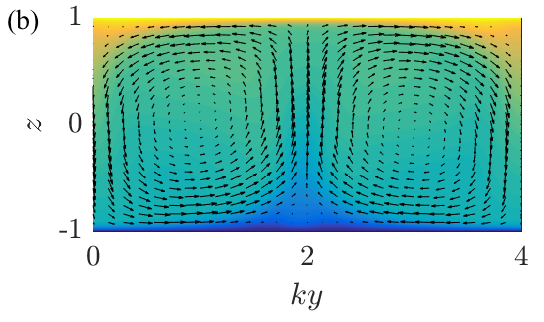}
\caption{(a) A roll cell in R2. 
(b) A roll cell in R3.
Legends are the same as in figure \ref{fig:xia-R23}.
The $y$ axis is scaled such that $ky$ is from 0 to $4$.
Again, the exact value of the contour is not relevant.
}
\label{fig:dans}
\end{figure}

The DNS solution gives a good ansatz for the roll cells.
Figures \ref{fig:dans} (a, b) show the re-scaled roll cells in R2 and R3, and they are very similar.
The following calculation will use the roll-cell solution in R2.

The velocities $u$, $v$, $w$ in R2 are related to the ansatz as follows
\begin{equation}
u\sim \beta,~~~
v\sim \partial_z\alpha,~~~
w\sim -(2/\pi)\partial_{ky}\alpha.
\end{equation}
%which allows us to numerically compute $\alpha$ from $v$ and $w$.

Repeating step five and plugging $\alpha$ and $\beta$ into  \eqref{eq:u1-3} and \eqref{eq:w1-3} and projecting  \eqref{eq:u1-3} and \eqref{eq:w1-3} on $\alpha$ and $\beta$, we have
\begin{equation}
\begin{split}
    \frac{db}{dt}&=(0.15+0.85~Ro)k~a-{(7.4k^2+14)}\frac{b}{R}\\
    \frac{da}{dt}&=-\frac{1}{R}(2.4k^2+2.4)-\frac{0.31k}{k^2+1}Ro~b.
\end{split}
\label{eq:data-ab}
\end{equation}
Taking fourth order derivative, e.g., $\partial_2^4$, $\partial_3^4$, lends the solution near the domain boundaries to significant numerical errors.
To avoid such errors, all integrations will be limited to wall-normal regions outside the viscous sublayer.
In arriving at \eqref{eq:data-ab}, we have also neglected terms that are significantly smaller than the retained ones. 

\begin{figure}
\centering
\vspace{2mm}
\includegraphics[width=0.45\textwidth]{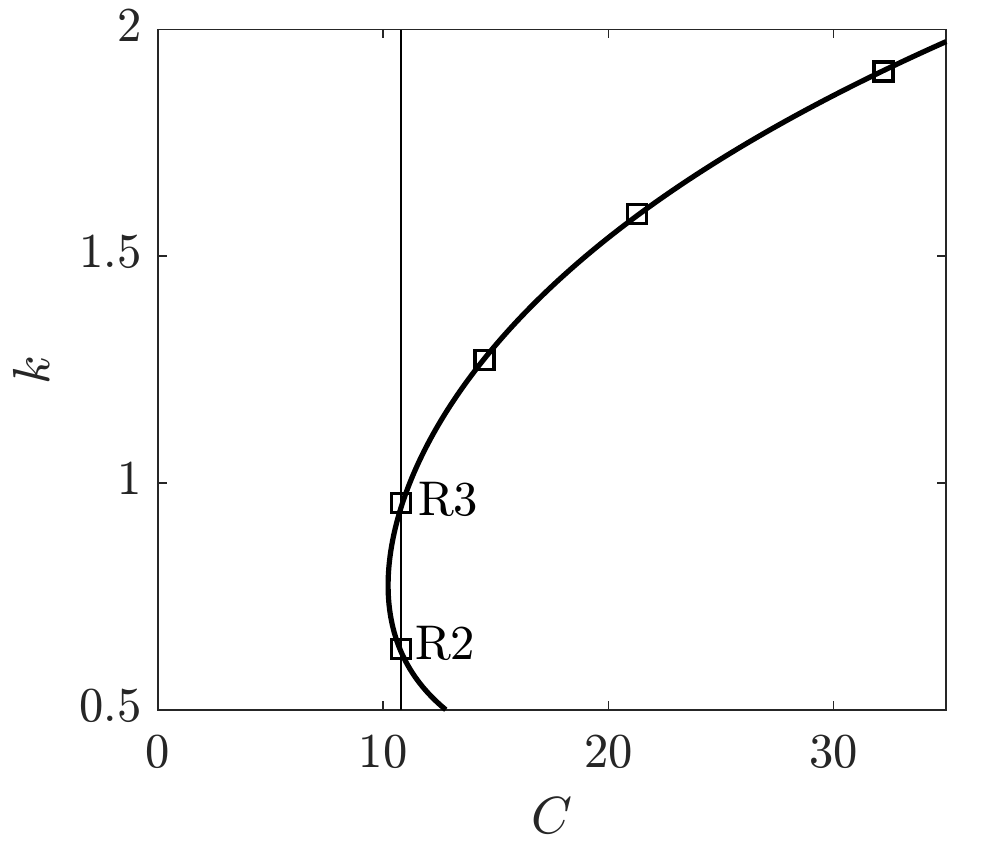}
\caption{Bifurcation diagram of the system in  \eqref{eq:data-ab}.
}
\label{fig:k-dans}
\end{figure}

The presence of multiple states requires \eqref{eq:data-ab} to have two or more non-trivial nodes, which in turn requires the following cubic equation of $k^2$ have two physically viable roots:
\begin{equation}
    {(k^2+1)^2(k^2+14/7.8)}-Ck^2=0.
    \label{eq:k2}
\end{equation}
Figure \ref{fig:k-dans} shows the bifurcation diagram, which leads to the same conclusions.

\section{Shell model}
\label{app:SM}

\begin{figure}
\centering
\vspace{2mm}
\includegraphics[width=0.45\textwidth]{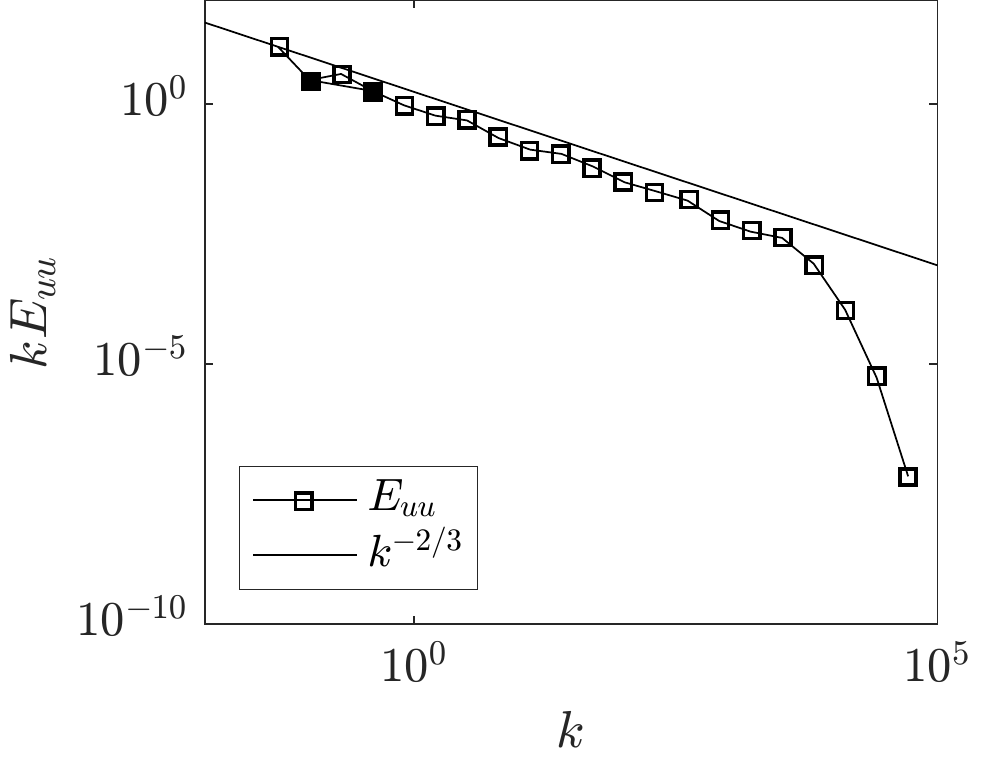}
\caption{Premultiplied shell model energy spectrum.
The 2nd and the 4th modes, whose trajectory is plotted in figure \ref{fig:phase-comp}, are colored black.
}
\label{fig:SM-Euu}
\end{figure}

The shell model of energy cascades in an isotripic turbulent flow reads
\begin{equation}
    \left(\frac{d}{dt}+\nu k^2_n\right)u_n=i\left(k_nu_{n+2}^*u_{n+1}^*-bk_{n-1}u_{n+1}^*u_{n-1}^*-ck_{n-2}u_{n-1}^*u_{n-2}^*\right)+f_n.
\end{equation}
It models the time evolution of a velocity fluctuation $u_n(t)$ over a wavelength $k_n=k_0\lambda^n$, with $\lambda$, the intershell ratio, usually set to 2.
The nonlinear coupling conserves the energy with $b=1/2$ and $c=-(1-b)$.
A equilibrium condition is obtained by forcing the flow at the large scale, i.e., $f_n=\delta_{1n}$.
Despite its simple form, shell model yields flow statistics that are nearly identical to those from an isotropic turbulent flow.
Further details of the shell model could be found in the review by \cite{biferale2003shell}. 
The trajectory in figure \ref{fig:phase-comp} (b) is a plot in the phase space of $\left|u_2\right|$ and $\left|u_4\right|$.
For our calculation, the number of shells is $N=21$, the wavenumber of the first shell is $k_1=0.05$, the molecular viscosity is $\nu=5e-6$, and the forcing term is $f_1=0.1(1+1j)$.
Figure \ref{fig:SM-Euu} shows the premultiplied shell model energy spectrum, $kE_{uu}$.
We see that $kE_{uu}$ follows $k^{-2/3}$ closely.

%%%%%%%%%%%%%%%%%%%%%%%%%%%%%%%%%%%%%%%%%%%%%%%%%%%%%%%%%%%%%%%%%%%%%%%%%%%%%%%%%%%%%%%%%
%\renewcommand{\thepage}{\arabic{page}}
\bibliographystyle{abbrvnat}
\bibliography{a-ref}

\begin{thebibliography}{34}
\providecommand{\natexlab}[1]{#1}
\providecommand{\url}[1]{\texttt{#1}}
\expandafter\ifx\csname urlstyle\endcsname\relax
  \providecommand{\doi}[1]{doi: #1}\else
  \providecommand{\doi}{doi: \begingroup \urlstyle{rm}\Url}\fi

\bibitem[Ahlers et~al.(2011)Ahlers, Funfschilling, and
  Bodenschatz]{ahlers2011heat}
G.~Ahlers, D.~Funfschilling, and E.~Bodenschatz.
\newblock Heat transport in turbulent rayleigh-b{\'e}nard convection for
  {P}r$\approx 0.8$ and {R}a$\lesssim 10^{15}$.
\newblock In \emph{Journal of Physics: Conference Series}, volume 318, page
  082001. IOP Publishing, 2011.

\bibitem[Anderson(2019)]{anderson2019non}
W.~Anderson.
\newblock Non-periodic phase-space trajectories of roughness-driven secondary
  flows in high-{R}e boundary layers and channels.
\newblock \emph{J. Fluid Mech.}, 869:\penalty0 27--84, 2019.

\bibitem[Biferale(2003)]{biferale2003shell}
L.~Biferale.
\newblock Shell models of energy cascade in turbulence.
\newblock \emph{Ann. Rev. Fluid Mech.}, 35\penalty0 (1):\penalty0 441--468,
  2003.

\bibitem[Drazin and Reid(2004)]{drazin2004hydrodynamic}
P.~G. Drazin and W.~H. Reid.
\newblock \emph{Hydrodynamic stability}.
\newblock Cambridge University Press, 2004.

\bibitem[Farrell and Ioannou(2007)]{farrell2007structure}
B.~F. Farrell and P.~J. Ioannou.
\newblock Structure and spacing of jets in barotropic turbulence.
\newblock \emph{Journal of the atmospheric sciences}, 64\penalty0
  (10):\penalty0 3652--3665, 2007.

\bibitem[Farrell and Ioannou(2014)]{farrell2014statistical}
B.~F. Farrell and P.~J. Ioannou.
\newblock Statistical state dynamics: a new perspective on turbulence in shear
  flow., 2014.

\bibitem[Farrell et~al.(2017)Farrell, Gayme, and
  Ioannou]{farrell2017statistical}
B.~F. Farrell, D.~F. Gayme, and P.~J. Ioannou.
\newblock A statistical state dynamics approach to wall turbulence.
\newblock \emph{PHILOS T R SOC A}, 375\penalty0 (2089):\penalty0 20160081,
  2017.

\bibitem[Frisch(1995)]{frisch1995turbulence}
U.~Frisch.
\newblock \emph{Turbulence: the legacy of AN Kolmogorov}.
\newblock Cambridge university press, 1995.

\bibitem[Galanti and Tsinober(2004)]{galanti2004turbulence}
B.~Galanti and A.~Tsinober.
\newblock Is turbulence ergodic?
\newblock \emph{Physics Letters A}, 330\penalty0 (3-4):\penalty0 173--180,
  2004.

\bibitem[Gul et~al.(2018)Gul, Elsinga, and Westerweel]{gul_2018}
M.~Gul, G.~E. Elsinga, and J.~Westerweel.
\newblock Experimental investigation of torque hysteresis behaviour of
  {Taylor--Couette} flow.
\newblock \emph{J. Fluid Mech.}, 836:\penalty0 635--648, 2018.

\bibitem[Huang et~al.(2019)Huang, Xia, Wan, Shi, and Chen]{huang2019hysteresis}
Y.~Huang, Z.~Xia, M.~Wan, Y.~Shi, and S.~Chen.
\newblock Hysteresis behavior in spanwise rotating plane couette flow with
  varying rotation rates.
\newblock \emph{Phys. Rev. Fluids}, 4\penalty0 (5):\penalty0 052401, 2019.

\bibitem[Huisman et~al.(2014)Huisman, Van Der~Veen, Sun, and
  Lohse]{huisman2014multiple}
S.~G. Huisman, R.~C. Van Der~Veen, C.~Sun, and D.~Lohse.
\newblock Multiple states in highly turbulent taylor--couette flow.
\newblock \emph{Nature communications}, 5\penalty0 (1):\penalty0 1--5, 2014.

\bibitem[Majda and Timofeyev(2000)]{majda2000remarkable}
A.~J. Majda and I.~Timofeyev.
\newblock Remarkable statistical behavior for truncated {B}urgers--{H}opf
  dynamics.
\newblock \emph{Proc Natl Acad Sci}, 97\penalty0 (23):\penalty0 12413--12417,
  2000.

\bibitem[Marusic et~al.(2013)Marusic, Monty, Hultmark, and
  Smits]{marusic2013logarithmic}
I.~Marusic, J.~P. Monty, M.~Hultmark, and A.~J. Smits.
\newblock On the logarithmic region in wall turbulence.
\newblock \emph{J. Fluid Mech.}, 716, 2013.

\bibitem[Rap{\'u}n and Vega(2010)]{rapun2010reduced}
M.-L. Rap{\'u}n and J.~M. Vega.
\newblock Reduced order models based on local {POD} plus {G}alerkin projection.
\newblock \emph{J Comput Phys}, 229\penalty0 (8):\penalty0 3046--3063, 2010.

\bibitem[Ravelet et~al.(2004)Ravelet, Mari{\'e}, Chiffaudel, and
  Daviaud]{ravelet2004multistability}
F.~Ravelet, L.~Mari{\'e}, A.~Chiffaudel, and F.~Daviaud.
\newblock Multistability and memory effect in a highly turbulent flow:
  Experimental evidence for a global bifurcation.
\newblock \emph{Phys. Rev. Lett.}, 93\penalty0 (16):\penalty0 164501, 2004.

\bibitem[Ravelet et~al.(2008)Ravelet, Chiffaudel, and
  Daviaud]{ravelet2008supercritical}
F.~Ravelet, A.~Chiffaudel, and F.~Daviaud.
\newblock Supercritical transition to turbulence in an inertially driven von
  k{\'a}rm{\'a}n closed flow.
\newblock \emph{J. Fluid Mech.}, 601:\penalty0 339--364, 2008.

\bibitem[Stevens et~al.(2009)Stevens, Zhong, Clercx, Ahlers, and
  Lohse]{Stevens2009}
R.~Stevens, J.~Zhong, H.~Clercx, G.~Ahlers, and D.~Lohse.
\newblock {Transitions between turbulent states in rotating Rayleigh--B\'enard
  Convection}.
\newblock \emph{Phys.~Rev.~Lett.}, 103:\penalty0 024503, 2009.

\bibitem[Taira et~al.(2017)Taira, Brunton, Dawson, Rowley, Colonius, McKeon,
  Schmidt, Gordeyev, Theofilis, and Ukeiley]{taira2017modal}
K.~Taira, S.~L. Brunton, S.~T. Dawson, C.~W. Rowley, T.~Colonius, B.~J. McKeon,
  O.~T. Schmidt, S.~Gordeyev, V.~Theofilis, and L.~S. Ukeiley.
\newblock Modal analysis of fluid flows: An overview.
\newblock \emph{AIAA Journal}, pages 4013--4041, 2017.

\bibitem[Tsinober(2001)]{tsinober2001informal}
A.~Tsinober.
\newblock \emph{An informal introduction to turbulence}, volume~63.
\newblock Springer Science \& Business Media, 2001.

\bibitem[van~der Poel et~al.(2011)van~der Poel, Stevens, and
  Lohse]{vanderpoel2011}
E.~van~der Poel, R.~Stevens, and D.~Lohse.
\newblock {Connecting flow structures and heat flux in turbulent
  Rayleigh--B\'enard convection}.
\newblock \emph{Phys.~Rev.~E}, 84:\penalty0 045303, 2011.

\bibitem[van~der Veen et~al.(2016)van~der Veen, Huisman, Dung, Tang, Sun,
  Lohse, et~al.]{van2016exploring}
R.~C. van~der Veen, S.~G. Huisman, O.-Y. Dung, H.~L. Tang, C.~Sun, D.~Lohse,
  et~al.
\newblock Exploring the phase space of multiple states in highly turbulent
  taylor-couette flow.
\newblock \emph{Phys. Rev. Fluids}, 1\penalty0 (2):\penalty0 024401, 2016.

\bibitem[Wang et~al.(2018)Wang, Wan, Yan, and Sun]{WangQ2018}
Q.~Wang, Z.~Wan, R.~Yan, and D.~Sun.
\newblock {Multiple states and heat transfer in two-dimensional tilted
  convection with large aspect ratios}.
\newblock \emph{Phys.~Rev.~Fluids}, 3:\penalty0 113503, 2018.

\bibitem[Wei et~al.(2015)Wei, Weiss, and Ahlers]{Wei2015}
P.~Wei, S.~Weiss, and G.~Ahlers.
\newblock Multiple transitions in rotating turbulent {Rayleigh--B\'enard
  Convection}.
\newblock \emph{Phys.~Rev.~Lett.}, 114:\penalty0 114506, 2015.

\bibitem[Weiss and Ahlers(2013)]{weiss2013effect}
S.~Weiss and G.~Ahlers.
\newblock Effect of tilting on turbulent convection: cylindrical samples with
  aspect ratio $\gamma= 0.50$.
\newblock \emph{J. Fluid Mech.}, 715:\penalty0 314--334, 2013.

\bibitem[Weiss et~al.(2010)Weiss, Stevens, Zhong, Clercx, Lohse, and
  Ahlers]{Weiss2010}
S.~Weiss, R.~Stevens, J.~Zhong, H.~Clercx, D.~Lohse, and G.~Ahlers.
\newblock Finite-size effects lead to supercritical bifurcations in turbulent
  {rotating Rayleigh--B\'enard Convection}.
\newblock \emph{Phys.~Rev.~Lett.}, 105:\penalty0 224501, 2010.

\bibitem[Xi and Xia(2008)]{xi2008flow}
H.-D. Xi and K.-Q. Xia.
\newblock Flow mode transitions in turbulent thermal convection.
\newblock \emph{Phys. Fluids}, 20\penalty0 (5):\penalty0 055104, 2008.

\bibitem[Xia et~al.(2018)Xia, Shi, Cai, Wan, and Chen]{xia2018multiple}
Z.~Xia, Y.~Shi, Q.~Cai, M.~Wan, and S.~Chen.
\newblock Multiple states in turbulent plane couette flow with spanwise
  rotation.
\newblock \emph{J. Fluid Mech.}, 837:\penalty0 477--490, 2018.

\bibitem[Xia et~al.(2019)Xia, Shi, Wan, Sun, Cai, and Chen]{Xia2019-analysis}
Z.~Xia, Y.~Shi, M.~Wan, C.~Sun, Q.~Cai, and S.~Chen.
\newblock Role of the large-scale structures in spanwise rotating plane couette
  flow with multiple states.
\newblock \emph{Phys. Rev. Fluids}, 4:\penalty0 104606, 2019.

\bibitem[Xie et~al.(2018)Xie, Ding, and Xia]{xie2018}
Y.-C. Xie, G.-Y. Ding, and K.~Xia.
\newblock Flow topology transition via global bifurcation in thermally driven
  turbulence.
\newblock \emph{Phys.~Rev.~Lett.}, 120:\penalty0 214501, 2018.

\bibitem[Yang et~al.(2020{\natexlab{a}})Yang, Xia, Lee, Lv, and
  J]{yang2020mean}
X.~I.~A. Yang, Z.~H. Xia, J.~Lee, Y.~Lv, and Y.~J.
\newblock Mean flow scaling in a spanwise rotating channel.
\newblock \emph{Phys. Rev. Fluids}, 2020{\natexlab{a}}.
\newblock \doi{In press}.

\bibitem[Yang et~al.(2020{\natexlab{b}})Yang, Chen, Verzicco, and
  Lohse]{YangYT-DDC}
Y.~Yang, W.~Chen, R.~Verzicco, and D.~Lohse.
\newblock Multiple states and transport properties of double-diffusive
  convection turbulence.
\newblock \emph{Proc. Nat. Acad. Sci.}, 2020{\natexlab{b}}.
\newblock \doi{10.1073/pnas.2005669117}.

\bibitem[Yokoyama and Takaoka(2017)]{Yokoyama-PRF}
N.~Yokoyama and M.~Takaoka.
\newblock Hysteretic transitions between quasi-two-dimensional flow and
  three-dimensional flow in forced rotating turbulence.
\newblock \emph{Phys. Rev. Fluids}, 2:\penalty0 092602(R), 2017.

\bibitem[Zimmerman et~al.(2011)Zimmerman, Triana, and Lathrop]{zimmerman2011bi}
D.~S. Zimmerman, S.~A. Triana, and D.~P. Lathrop.
\newblock Bi-stability in turbulent, rotating spherical couette flow.
\newblock \emph{Phys. Fluids}, 23\penalty0 (6):\penalty0 065104, 2011.

\end{thebibliography}
%}

\end{document}